\documentclass{aa}  

\usepackage{mathrsfs}
\usepackage{url}
\usepackage{hyperref}
\hypersetup{
    colorlinks=true,
    linkcolor=blue,
    filecolor=black,      
    urlcolor=blue,
    citecolor=blue,
    pdftitle={Overleaf Example},
    pdfpagemode=FullScreen,
    }
\usepackage{comment}

\usepackage{graphicx}
\usepackage{txfonts}
\usepackage{lipsum}
\usepackage{subcaption}  
\usepackage{ulem}
\usepackage{xcolor}
\usepackage{lscape}             
\usepackage{placeins}     

\renewcommand{\linenumbers}{}

\begin{document}

   \title{A systematic search for orbital periods of polars with TESS\thanks{Table~\ref{table:Master} is available at the CDS via  anonymous ftp to \href{http://cdsarc.u-strasbg.fr/}{cdsarc.u-strasbg.fr} (\href{ftp://130.79.128.5/}{130.79.128.5}) or via 
\url{http://cdsweb.u-strasbg.fr/cgi-bin/qcat?J/A+A/}}}

   \subtitle{Methods, detection limits, and results}

   \author{S. Hernández-Díaz\inst{1}
        \and B. Stelzer\inst{1} \and A. Schwope\inst{2} \and D. Muñoz-Giraldo\inst{1}
        }

\institute{Institut für Astronomie und Astrophysik, Eberhard Karls Universität Tübingen, Sand 1, 72076 Tübingen, Germany\\
\email{hernandez@astro.uni-tuebingen.de}
\and Leibniz Institut für Astrophysik Potsdam (AIP), An der Sternwarte 16, 14482 Potsdam, Germany\\ }
\date{Received XX / Accepted XX}

 
\abstract
   {Determining the orbital periods of cataclysmic variable stars (CVs) is essential for confirming candidates and for the understanding of their evolutionary state. The Transiting Exoplanet Survey Satellite (TESS) provides month-long photometric data across nearly the entire sky that can be used to search for periodic variability in such systems.}
   {This study aims to identify and confirm the orbital periods for members of a recent compilation of magnetic CVs (known as polars) using TESS light curves. In addition to providing the periods, we set out to investigate their reliability, and hence the relevance of TESS for variability studies of CVs.} 
   {Four period-search methods were used, namely the Lomb-Scargle periodogram, the autocorrelation function (ACF), sine fitting, and Fourier power spectrum analysis, to detect periodic signals in TESS light curves. We investigated the correlation between noise level and TESS magnitude by 'flattening' the observed TESS light curves, effectively isolating the noise from the periodic modulation. To evaluate the reliability of the period detections, we developed a probabilistic framework for the detection success across signal-to-noise ratios (S/N) in the power spectral density of observed light curves.}
   {Ninety-five of the $217$ polars in our sample have pipeline-produced TESS two-minute cadence light curves available. The results from our period search are overall in good agreement with the previously reported values. Out of the 95 analysed systems, 85 exhibit periods consistent with the literature values. Among the remaining ten objects, four are asynchronous polars, where TESS light curves resolve the orbital period, the white dwarf’s spin period, and additional beat frequencies. For four systems, the periods detected from the TESS data differ from those previously reported. For two systems, a period detection was not possible due to the high noise levels in their light curves. 
   Our analysis of the flattened TESS light curves reveals a positive correlation between noise levels—expressed as the standard deviation of the flattened light curve—and TESS magnitude. Our noise level estimates resemble the rmsCDPP, a measure of white noise provided with the TESS pipeline products. However, our values for the noise level are systematically higher than the rmsCDPP indicating red noise and high-frequency signals hidden in the flattened light curves. 
   Additionally, we present a statistical methodology to assess the reliability of period detections in TESS light curves. We find that for TESS magnitudes\,$\gtrsim$17, period detections become increasingly unreliable.
   }
   {Our study shows that TESS data can be used to reliably and efficiently determine orbital periods in CVs. The developed methodology for period detection, noise characterisation, and reliability assessment can be systematically applied to other variable star studies, thus improving the robustness of period measurements in large photometric data sets.}


   \maketitle

\section{Introduction}

Cataclysmic variable stars (CVs) are a class of close binary star systems consisting of a white dwarf (WD)  primary and a companion star, typically a  low-mass main-sequence star. These systems exhibit extreme variability due to mass transfer from the secondary star (the donor) to the WD. This mass transfer occurs when the secondary component fills its Roche lobe, the region within which material remains gravitationally bound, allowing material to overflow onto the white dwarf (see \citealt{Warner_Book} for a comprehensive review). 

Generally, CVs can be classified based on the presence of a WD’s magnetic field in non-magnetic and magnetic systems, with the latter subdivided into intermediate polars and polars according to the strength of the WD’s magnetic field. 
The presence or absence of a strong magnetic field of the WD is a fundamental property of the system, since it changes the nature of the accretion process and the system’s observable features. In systems without a magnetic field, accretion occurs through the formation of an accretion disk. In contrast, in polars the strong magnetic field of the WD ($B_{\rm WD} > 10^{6}$ G) \citep{Cropper_Polars} prevents the formation of an accretion disk. Instead, the material overflowing the Roche lobe of the secondary is funnelled along the field lines and accretes directly onto the magnetic poles of the WD, producing extreme ultraviolet and optical emission from the accretion hotspots of the WD. In intermediate polars, the weaker WD's magnetic field ($B_{\rm WD} \sim 10^{6}$ G) \citep{Patterson_1994} truncates the accretion disk at its inner parts, leading to the formation of a partial disk and magnetically controlled accretion in the truncated regions (see \citealt{Cropper_Polars} and \citealt{Patterson_1994} for a comprehensive review). 

In polars, the WD rotation is synchronised with the orbital motion due to the dipole-dipole magnetostatic interaction between the magnetic fields of the primary and the secondary (see \citealt{MHD_Torque}, \citealt{Campbell_1} and \citealt{Campbell_2}). As a consequence, the hotspots co-rotate with the WD, resulting in  variability in polars that is modulated at the orbital (alias WD spin) period. 
 
While most polars exhibit strict synchronisation, a small subset known as asynchronous polars deviates from this behaviour. In asynchronous polars, the spin period of the white dwarf differs from the orbital period by a few percent. This state is thought to originate from recent nova explosions (see e.g. \citealt{Nova_Asynchronous_Polars}) and to be transient;   these systems  gradually evolve toward synchronisation, a process that can take between 100 to 6000 years (see e.g. \citealt{Campbell_Schwope_1999}, \citealt{Honeycutt&Kafka}, and \citealt{Resynchronization}). Studying the optical photometric variability of polars allows  precise measurements of their periods. 

Measuring orbital periods of CVs is of crucial importance for the confirmation of candidates, as periodic variability serves as direct evidence of accretion and hence of their binary nature. For this article, we conducted a systematic search for orbital periods in known polars based on observations from the Transiting Exoplanet Survey Satellite (TESS;  \citealt{TESS}). 
Although TESS was designed to search for exoplanets, its month-long light curves make it exceptionally well suited for the measurement of orbital periods of CVs. Several recent studies have made use of TESS light curves for measuring the periodic variability of CVs. A series of five articles by Albert Bruch (2022–2024) (\citealt{Bruch_1}, \citealt{Bruch_2}, \citealt{Bruch_3}, \citealt{Bruch_4}, and \citealt{Bruch_5}) presented comprehensive sample studies of nova-like CVs and old novae, leading to the discovery of new orbital periods and superhump signals. These studies also refined and corrected previously reported orbital periods. Other studies have focused on individual systems, using TESS photometry to detect not only orbital periods but also spin periods and beat frequencies in asynchronous polars (see e.g. \citealt{Mason_2}, \citealt{49}, \citealt{Littlefield_2}, and \citealt{48}). These works highlight the potential of TESS light curves for period analysis in CVs.  However, CVs are typically  faint, with TESS magnitudes $T$ between $13$ and $19$ in our sample (see left panel in Fig.~\ref{Fig.MagHistogram}), while TESS was designed for observations of bright nearby stars with $T<13$ (\citealt{Tmag}). Therefore, special care needs to be attributed to an evaluation of the significance of period detections in CVs. Here, we introduce a systematic approach to  quantify the reliability of TESS-based periods by a study of the signal-to-noise ratio (S/N) of the power spectral density and how this depends on the TESS magnitude of the objects, which determines the noise level of the light curves. 

We describe the sample of polars in Sect.~\ref{sect.Catalog} and the TESS database in Sect.~\ref{sect.TESS_Data_Base}. Our methodology for period detection and the results from the period search are presented in Sect.~\ref{Search_Period}. In Sect.~\ref{sec.Flat_LCs} we establish the correlation between the noise level in the  TESS light curves and the TESS magnitude. Sect.~\ref{Reliability_Periods} introduces a  statistical methodology to assess the reliability of orbital period detections, from which we evaluate a limiting TESS magnitude for which orbital period detections in CVs  become unreliable. Finally, in Sect.~\ref{Summary} we provide a summary  of our study and draw our conclusions.

\section{Master catalogue of polars}
\label{sect.Catalog}

This article is based on a comprehensive catalogue of polars that we compiled from the literature. The full catalogue of polars is the subject of a separate publication  (\citealt{PolarCat}). The list of polars continuously evolves, and we used the preliminary version of the catalogue compiled up to Summer 2024. This catalogue comprises $217$ systems. It serves as the input list for our systematic search for rotation periods in TESS data. 

The catalogue of polars comprises orbital periods that were obtained from a meticulous  literature study and which serve in the context of this work to  validate  the periods that we detect with TESS.

We adopted the \textit{Gaia}\,DR3 source IDs (\citealt{Gaia_DR3}), hereafter \textit{Gaia}\,DR3 IDs, provided by \citealt{PolarCat}, for all systems except for 2XMM\,J154305.5-522709. For this system, a TESS two-minute cadence light curve is available (see  Sect.~\ref{sect.TESS_Data_Base}), but no \textit{Gaia}\,DR3 ID is given in \citealt{PolarCat}. Therefore, we used  the J2000 equatorial coordinates to perform a cross-match in TOPCAT (\citealt{Topcat}) and retrieve its \textit{Gaia}\,DR3 ID. For all systems, the \textit{Gaia}\,DR3 photometry was subsequently acquired through the \textit{Gaia} Archive.\footnote{\citealt{gaia_archive}}

\section{TESS database}
\label{sect.TESS_Data_Base}

TESS is a NASA space-based mission led by MIT. Initially designed to conduct an almost full-sky survey (excluding the ecliptic plane) over two years, the mission has been extended and continues to re-observe the sky, now entering its additional sixth year beyond the original timeline. The observational strategy of TESS divides the sky into sectors, each covered for two consecutive satellite orbits ($\sim$ 27 days). In the mission's first year (2018-2019), the southern ecliptic hemisphere was observed, followed by the northern ecliptic hemisphere in the second year (2019-2020). In its extended mission, TESS continues its sector-based sky survey with expanded target lists and additional coverage of the ecliptic plane. As of February 2025, TESS has completed observations through sector\,89.

In this work we use data from Sectors~1 to 63. First, we performed a cross-match within TOPCAT using the J2000  equatorial coordinates. This way, we retrieved the unique identifiers of our targets in the TESS Input Catalogue (\citealt{TIC}), hereafter referred to as TIC IDs, and their TESS magnitudes. Then we searched in the Barbara A. Mikulski Archive for Space Telescopes (MAST)\footnote{\citealt{mast_portal}} for two-minute cadence data of the $217$ polars from our catalogue. We found that $95$ of them have two-minute cadence data from at least one of the sectors (within  Sectors~1 to 63).  
Many of our targets have several TESS two-minute cadence light curves from sectors corresponding to different years of observation or from sectors observed directly one after each other if the star is in the overlapping sky area of the two sectors. For each system in our sample we analysed all two-minute cadence light curves available from Sector\,1-63.  

\begin{figure*}
    \centering
    \includegraphics[width=0.9\linewidth]{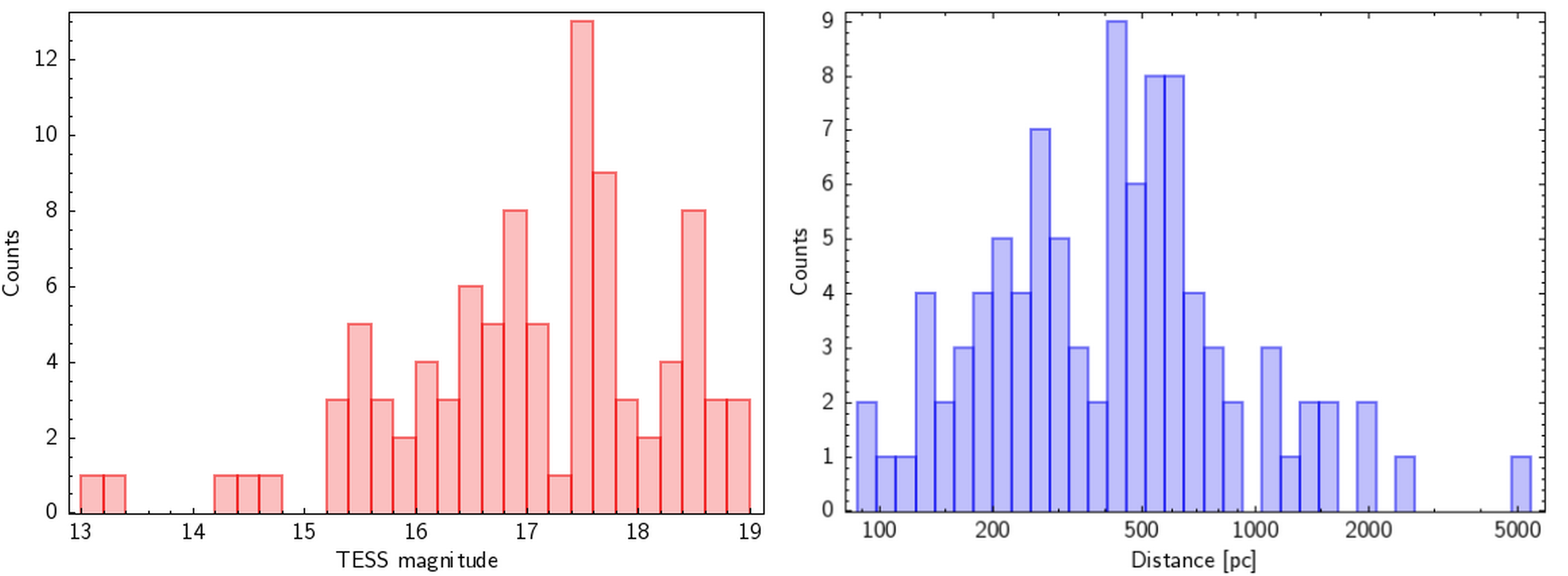}
    \caption{Properties of the sample of polars with two-minute time-resolution TESS light curves. Left panel: Histogram of TESS magnitudes. Right panel: Histogram of distances (from \citealt{Bailer-Jones}).}
    \label{Fig.MagHistogram}
\end{figure*}

The distributions of TESS magnitudes and distances from \cite{Bailer-Jones} for our sample are shown in the left and right panels of Fig.~\ref{Fig.MagHistogram}, respectively.  We analysed the light curves with the PreSearch Data Conditioning Simple Aperture Photometry (PDCSAP) fluxes, which is the pipeline-corrected version of the photometric fluxes that mitigates instrumental effects. The journal of observations, combined with the data table for our results from the period analysis, is available electronically at CDS (see Table~\ref{table:Master} in Appendix~\ref{sec:Table} for details about the content).

\section{Search for orbital periods}
\label{Search_Period}

\subsection{Data preprocessing}

As a first step, we removed  data points with TESS quality flags different from zero, ensuring that unreliable or flagged measurements are excluded from the analysis.

TESS light curves contain gaps of $\sim1-2$ days, the Low Altitude Housekeeping Operations (LAHO) gaps, used for data transfer to Earth or, occasionally, due to technical problems. These  discontinuities are potentially problematic for our  analysis, impeding some algorithms to work properly. Therefore, we performed a linear interpolation across all gaps—both those associated with LAHO and missing points associated with problematic quality flags—between the last observed point before each gap and the first observed point after it.

To assess whether the linear interpolation biases the period estimation, we compared the period estimates using the Lomb-Scargle periodogram on our sample with and without applying linear interpolation. The results showed a mean percentage difference of 0.020\% and a maximum difference of 0.098\%, both values within the typical noise uncertainties. Thus, we conclude that the linear interpolation does not significantly bias the period determination.

Nevertheless, the interpolated light curves were only used for the autocorrelation function (ACF) and Fourier power spectrum methods, for which uninterrupted data is required. No interpolation was applied when using the Lomb-Scargle periodogram or sine fitting methods (see Sect.~\ref{subsect:periods_methods}).

Finally, we normalised the light curves by the mean value of the photometric flux.

\subsection{Detection methods}
\label{subsect:periods_methods}

For the detection of orbital periods from TESS light curves, four numerical methods are employed: Lomb-Scargle periodogram, the autocorrelation function, sine fitting, and Fourier power spectrum analysis. Utilising multiple methods ensures robust and accurate period determination by mitigating biases inherent to individual techniques, uncovering subtle periodic signals that could be missed by individual methods alone and allowing cross-validation of the different period estimations.

The Lomb–Scargle periodogram (\citealt{Lomb}, \citealt{Scargle}) is an algorithm especially suitable for detecting periodic signals in unevenly sampled time series. 
The autocorrelation function measures the similarity between the observed light curve and a delayed version of itself revealing periodic patterns. The autocorrelation coefficient is computed at different time lags. A fast Fourier transform (FFT) (\citealt{FFT}) is then applied to the obtained autocorrelation coefficients, allowing the identification of the periodic signals that are present in the light curve. Sine fitting is performed using non-linear least squares using the Levenberg-Marquardt algorithm (\citealt{Levenberg}).
 Finally, a Fourier power spectrum with Kaiser windowing (\citealt{Kaiser_Window}) is obtained for each light curve using a beta parameter $\beta = 3$. The Kaiser window provides an important flexibility in trading off the main lobe width and the sidelobe levels by adjusting the beta parameter. With $\beta = 3$, the Kaiser window achieves a good compromise between suppressing sidelobes sufficiently and maintaining a good frequency resolution. 
 The software routines used in this work are based on Python implementations available in the \textsl{statsmodels}, \textsl{Astropy}, and \textsl{Scipy} packages.

\subsection{Results}\label{subsect:periods_results}
 
With the analysis described in Sect.~\ref{subsect:periods_methods} we obtain one period measurement per method per light curve, and given that some systems were observed in two-minute cadence throughout $N_{\rm sec}$ different sectors, the number of period values we obtain for a given star is up to $4\times N_{\rm sec}$. To determine the final adopted period for each system, we first compute sector-averaged periods for each period search method, that is, $\langle \text{P}_{\rm orb, LS} \rangle, \langle \text{P}_{\rm orb, ACF} \rangle, \langle \text{P}_{\rm orb, SINE} \rangle, \langle \text{P}_{\rm orb, FS} \rangle$; see description in Sect.~\ref{subsect:period_adopted}. This approach allows us to evaluate the performance of the different period search methods and to discard possible systematic effects. However, no systematic trends between the period values found with the different methods are observed. 

The use of multiple period detection methods and the availability of several light curves per star require to set up a procedure for the  determination of the final adopted orbital period, $\text{P}_{\rm orb,final}$, of each star. We explain our approach in Sects.~\ref{sec.Uncertainties}-\ref{subsect:period_adopted}, including the treatment of the uncertainties, the correction of modulations at $1/2\,\text{P}_{\text{orb}}$ and the procedure for determining the final adopted periods.

To validate our new period detections, we visually inspected the Lomb-Scargle periodogram, ACF, and Fourier power spectrum plots for each light curve.
Additionally, for each system in which a period could be measured from the TESS data, a comparison of $\text{P}_{\rm orb, final}$ against values previously reported in the literature was carried out and the measurements that deviate most from the earlier literature value for the same star were flagged.
Hereby, we use the most recent literature reference and value. We performed this cross-check by first calculating the difference between the literature period and the final adopted period, $\text{P}_{\rm orb,final}$, for each system, $\text{P}_{\text{dif}}=\text{P}_{\text{orb,lit}}- \text{P}_{\text{orb,final}}$. 
Then, we computed the 95th percentile of the absolute values of all $\text{P}{\text{dif}}$ values.  
All period measurements, $\text{P}_{\rm orb,final}$, for which $|\text{P}_{\text{dif}}|$ is smaller than this threshold were then deemed consistent with the literature value. 

Out of the 95 systems analysed, we were able to measure an orbital period for $93$ systems.  Among these, for $85$ systems the measured orbital periods are consistent with the values reported in the literature. The sample studied also  comprises four asynchronous systems which are analysed in detail in Sect.~\ref{Asynchronous}. For four systems, the measured orbital periods were not consistent with values reported in the literature. In these cases, a careful examination was performed that is explained in Sect.~\ref{Outliers}. Additionally, for two systems, the presence of noise in their TESS light curves impeded a period measurement (Sect.~\ref{Noisy_Cases}). 

In an electronic table available at CDS (see Table~\ref{table:Master} in Appendix~\ref{sec:Table}),  all the studied 95 polars are presented together with the orbital periods $\text{P}_{\rm orb,final}$ measured from their TESS two-minute cadence light curves and their uncertainties (explained in Sect.~\ref{sec.Uncertainties}).  We also provide the literature reference for the previously reported value of the orbital period $\text{P}_{\rm lit}$.

\subsection{Uncertainties}
\label{sec.Uncertainties}

Quantifying the uncertainties associated with the 
periods 
detected in light curves 
is notoriously challenging. In this context, common practice consists in measuring the width of the periodogram peak associated with the orbital period (see e.g. \citealt{Mennickent_2003} or \citealt{Nagel_2016}). 
The uncertainty is then directly ascribed to the half-width at half-maximum in the frequency  peak or the standard deviation obtained from a Gaussian fit to the frequency peak. However, this methodology is conceptually limited, as the width of the frequency peak is a naive estimation of the precision associated with a period measurement, ignoring factors such as the noise in the data. 
 
The reason is that these  uncertainty  estimates are intrinsically linked to the frequency resolution. In a Fourier transform, the frequency resolution is the inverse of the total observation time of the time-series data. In the discrete Fourier transform: $\Delta f = 1/(N\Delta t)$, where $N$ is the number of data points and $\Delta t$ is the time-resolution. Thus, given a fixed total observation time of the time-series data, the frequency resolution and, consequently, the width of the frequency peak in the periodogram, are, to a first order approximation, independent  
of both the sampling cadence and 
the S/N of the data points. In addition, the frequency width is highly dependent on how the periodogram is obtained. The use of different window functions can drastically change the obtained main lobe widths of the frequency peaks (see \citealt{LombScargleUnderstanding} for a detailed discussion).

Hence, this methodology  measures the precision  associated with the limited frequency resolution, which depends on the total observation time of the time-series data and not on the quality of the data, misconceiving the true variability in the detection process. Instead, it is more appropriate to estimate the uncertainty associated with the statistical variations, due to factors such as noise, instrumental systematics, and errors introduced by the period-search methods.

Thus, we made use of the multi-epoch data of our targets. Forty-three of the systems have more than one two-minute cadence TESS light curve. For these cases, we computed the uncertainty in the period on a per-method basis by calculating the 1-$\sigma$ confidence interval for the sample of periods retrieved with a given period-search method, as described in Sect.~\ref{subsect:period_adopted}. We note that this approach is generally applicable for polars, whose orbital periods remain stable over long timescales (typically decades or longer; see e.g. \citealt{Applegate&Patterson1987} and \citealt{Applegate1992}). For other types of variable stars, however, additional diagnostics may be necessary to ensure that the observed scatter in period estimates from different epochs is not due to evolutionary effects or episodic events like outbursts.

For the 52 systems with only one two-minute cadence TESS light curve, the uncertainties in the detected periods were estimated using a Monte Carlo resampling approach. Hereby, each Monte Carlo realisation of the observed light curve is produced by (1) adding noise drawn from a normal distribution with zero mean and a standard deviation equal to the photometric uncertainties of the original flux measurements, and (2) applying a resampling with replacement to the time–flux pairs, that is, time–flux pairs are randomly drawn from the original time–flux measurements, where replacement means that some time-flux pairs can be selected multiple times while others may be omitted. This procedure is repeated 1000 times and for each simulated light curve, the Lomb-Scargle periodogram is computed over a narrow frequency range centred on the originally detected frequency. Specifically, we used a frequency grid spanning ±10\% around the detected frequency peak. The distribution of the resulting dominant frequencies is cleaned using an iterative 3-$\sigma$ clipping procedure to remove outliers. In this process, values exceeding 3 standard deviations from the median are removed, after which the median and standard deviation are recalculated. This step is repeated until no further outliers are identified, leading to a Gaussian-like distribution. The frequency uncertainty is then estimated from the standard deviation of this distribution.

While this method captures the effects of random noise and sampling variability, we note that it may underestimate the full uncertainty associated with the period determination. For example, in each realisation of the observed light curve, it is assumed that the observational noise is purely white, that is uncorrelated, whereas in reality, different realisations of red noise can also affect the period determination. In addition, the LAHO gap in the observed TESS light curve is kept fixed in both position and duration across all Monte Carlo realisations, which neglects its potential impact in biasing the period determination. Finally, any instrumental systematics present in the original light curve are approximately preserved in all simulated realisations of the observed light curve, potentially biasing the resulting distribution of dominant frequencies.

\subsection{Periods at $1/2 \,\text{P}_{\text{orb}}$}
\label{sec.HalfPeriods}

During our search for orbital periods, dominant modulations in the light curves at half the period reported in the previous literature
were encountered  in 38 systems. If we assume that the literature values correspond to the orbital period, our measurements for these 38 systems represent half the orbital period ($1/2 \,\text{P}_{\rm orb}$). The detection of orbital periods in polars, indeed,  underlies  ambiguities because various physical processes may cause periodicities at one half of the orbital cycle. These phenomena are here briefly discussed before we explain how we mitigated wrong values of $\text{P}_{\rm orb}$ due to their presence.

\subsubsection{Physical mechanisms producing periods at $1/2\,\text{P}_{\rm orb}$}
\label{subsect:half_period_mechanism}

A primary cause of periodic light curve patterns at $1/2 \,\text{P}_{\text{orb}}$
is accretion to both magnetic poles of the white dwarf (see e.g. \citealt{Ferrario_2_poles} and \citealt{2_pole_accretion_example}). When accretion occurs at both magnetic poles, a distinct hotspot is produced at both polar regions. These spots  contribute  to the overall brightness of the system. Depending on the inclination angle, as the binary system orbits, these hotspots come in and out of view and give rise to a periodic modulation of the observed brightness. For appropriate inclinations the observer might see two maxima per orbital cycle, one from each magnetic pole, and a periodic signal corresponding to half the orbital period of the binary system will be detected.

The accreting region of the white dwarf can be the source of significant cyclotron emission due to the strong magnetic field of the white dwarf (see \citealt{Cyclotron_Radiation}). Cyclotron beaming combined with a misalignment between rotation and magnetic axis (that is a non-zero colatitude angle $\beta$) appears, thus, as another potential source of brightness modulations at $1/2 \,\text{P}_{\text{orb}}$ in light curves of polars. Cyclotron beaming is highly anisotropic and the emission is channelled along a direction approximately perpendicular to the magnetic axis. For $\beta \neq 0$ and inclination angle $i \neq 0$, the hotspot, located near the accreting magnetic pole, will produce cyclotron beaming that changes its orientation as the white dwarf rotates moving into and out of the observer's view. The variation in the orientation of the beamed radiation relative to the observer's line-of-sight leads to a periodic modulation of the observed brightness—with a typical amplitude of $\sim 1\,$mag (see e.g. \citealt{Vogel.etal_2007})—which has a period of  half the orbital period, since each complete rotation of the white dwarf results in two instances where the cyclotron beam is directed towards the observer.

Another possible cause of signals at $1/2 \,\text{P}_{\text{orb}}$ is ellipsoidal modulation. In CVs, the donor becomes distorted into an ellipsoid due to the strong gravitational influence of its close WD companion. This distortion causes periodic brightness variations—typically with amplitudes of $0.01-0.2\,$mag—as the projected surface area of the elongated star changes during the orbit (see e.g. \citealt{Ellipsoidal_Modulation} and \citealt{Half_Periods_1}). Generally, the same modulation pattern repeats twice per orbital period since the ellipsoid presents similar cross-sectional areas twice per orbit. This leads to a contribution to the light curve that exhibits two maxima and two minima per orbital period.

\subsubsection{Results for our sample}
\label{HalfPeriodsCorrection}

\begin{figure*}
	\centering
	\includegraphics[width=0.7\linewidth]{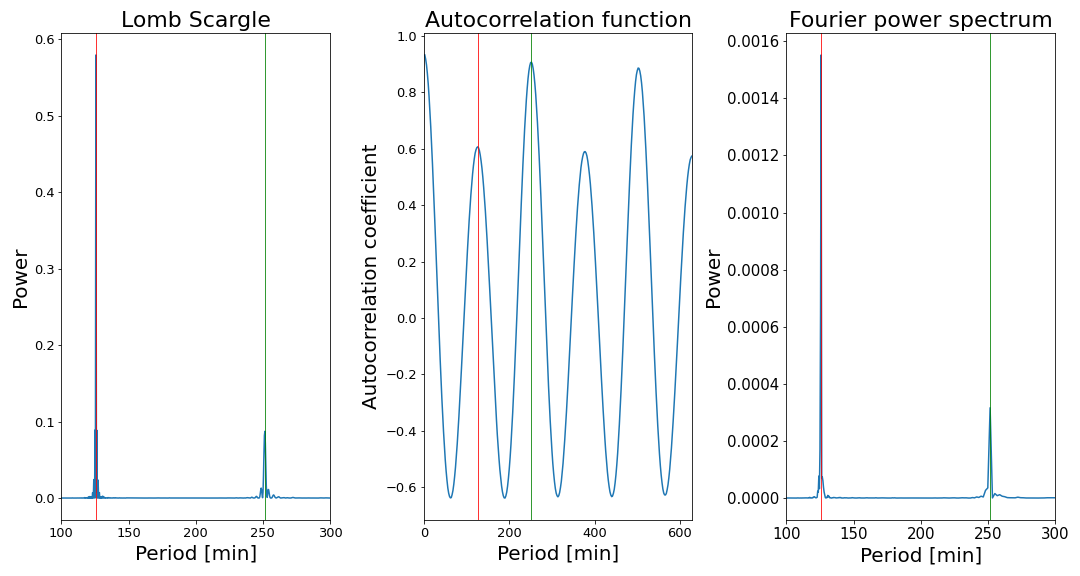}
\caption{Illustration of the results from different period detection methods for a case where the dominant period is $1/2 \,\text{P}_{\rm orb}$ (from left to right): Lomb-Scargle periodogram, ACF, and Fourier power spectrum. 
The example refers to  TIC\,124404442 (alias V1043 Cen) (sector\,37).  
The red vertical lines indicate the signals at $1/2 \,\text{P}_{\text{orb}}$, while the green vertical lines indicate the true orbital period (\citealt{4}). In this light curve, all three methods failed to identify the orbital period. We note that although the autocorrelation coefficient at the true orbital period is higher than that at $1/2 \,\text{P}_{\text{orb}}$ in the ACF plot, the subsequent FFT applied to the autocorrelation coefficients yields a stronger signal in $1/2 \,\text{P}_{\text{orb}}$.}
\label{Fig.Methods_Half_Period}
\end{figure*}

The identification of periods in the TESS light curves that represent  $1/2 \,\text{P}_{\text{orb}}$ was primarily done by comparison with previously reported values from the literature, which were taken as a reference for the orbital period of the systems. In addition, 
we searched for a signal at twice the measured value to double-check if the dominant observed period is, indeed, $1/2 \,\text{P}_{\text{orb}}$. If the detection that corresponds to the strongest signal in the light curve is half the orbital period, the signal corresponding to the actual orbital period should, in principle, also be present. Specifically, the Lomb-Scargle periodogram and the Fourier power spectrum should present two frequency peaks at the positions corresponding to $\text{P}_{\rm orb}$ and $1/2 \,\text{P}_{\rm orb}$.  
The ACF should exhibit two superposed periodic modulations. An example of such a case is illustrated in  Fig.~\ref{Fig.Methods_Half_Period}. 

\begin{figure}
	\centering
	\includegraphics[width=1.0\linewidth]{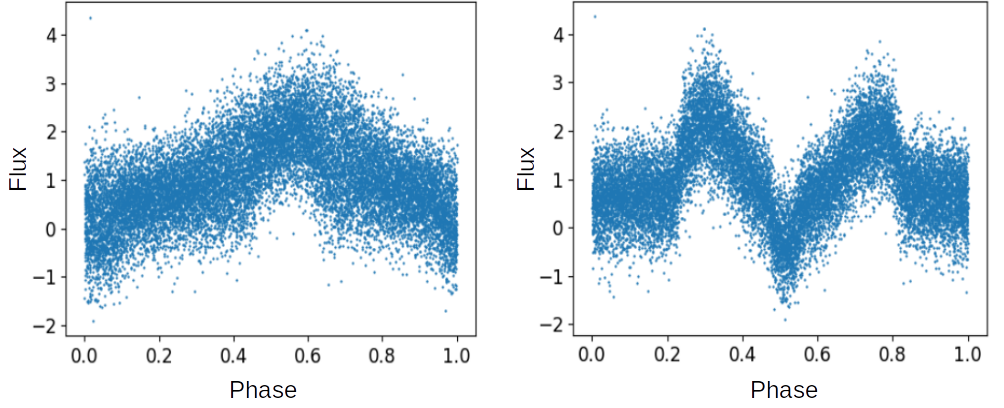}
	\caption{Phase plots for the light curve of TIC 258799357 (alias EP\,Dra) (sector 26), an eclipsing polar, as an example of a light curve where the dominating modulation is at $1/2 \,\text{P}_{\text{orb}}$. Left panel: Light curve folded over the detection at $1/2 \,\text{P}_{\text{orb}}$. Right panel: Light curve folded over the true orbital period \citep{41}. The true orbital period (right panel) displays the pattern corresponding to one complete orbit of the primary star around the secondary, with the eclipse distinctly visible around phase 0.5.}
	\label{Fig.Phase_Plot}
\end{figure}

The search for two periodic signals that differ by a factor two is particularly useful when applied to new samples of CV candidates without previously reported periods in the literature, providing a systematic way to verify whether the detected signal corresponds to half the true orbital period. Additionally, we recommend inspecting the phase-folded light curves for both the detected period and the possible true period at twice the period with the strongest signal. The light curve folded with $1/2 \,\text{P}_{\rm orb}$ shows the superposition of two different modulations, at $1/2 \,\text{P}_{\rm orb}$ and at $\text{P}_{\rm orb}$, typically with different amplitudes. The light curve folded with the true orbital period reveals these two modulations and aligns with other periodic phenomena, such as eclipses. An example illustrating these differences is shown in Fig.~\ref{Fig.Phase_Plot}.

Out of the $93$ systems for which a period was detected, we identified $38$ systems where some of the detections actually represent $1/2 \,\text{P}_{\rm orb}$. For some systems, different light curves yielded a different dominant periodicity, either $1/2 \,\text{P}_{\rm orb}$ or $\text{P}_{\rm orb}$. Moreover, in some instances, for a specific light curve, one numerical method identified $1/2 \,\text{P}_{\rm orb}$ as the strongest signal, while another method favoured $\text{P}_{\rm orb}$ as the dominant modulation. For all the  independent detections corresponding to $1/2 \,\text{P}_{\rm orb}$, the difference between the literature values and twice the  detected  periods is on the order of seconds or less, which is within the noise error. 
These values were, thus, corrected by simply taking the double of the detected period $\text{P}_{\rm det}$, that is the corrected period is $\text{P}_{\rm corr} := 2\text{P}_{\rm det}$. We note that, in general, doubling a noisy period estimate may misidentify $\text{P}_{\rm orb}$. However, in our study, no cases were encountered where the dominant periodicity at $1/2 \,\text{P}_{\rm orb}$ was noisy. In the table available at CDS (see Table~\ref{table:Master} in Appendix~\ref{sec:Table}), we provide a flag for the systems where at least one of the individual measurements  involved in calculating  the period we report has been corrected this way.

\subsection{Determination of final adopted orbital period values}
\label{subsect:period_adopted}

As mentioned in Sect.~\ref{subsect:periods_results}, since the same object may have been observed in $N_{\rm sec}$ different sectors, and for each light curve we measure the orbital period from four different numerical methods, for a given system we obtain up to $4\times N_{\rm sec}$ period values. We note that not for all light curves, a period was found with all detection methods. Conversely, for a given system, a particular detection method may not find a period in all the available light curves.  The following procedure was implemented to determine the final values for the orbital periods and their uncertainties.

First, all pathological detections were identified and discarded. This identification of erroneous detections was clear in many cases, where the detected periodicities were several orders of magnitude higher than the expected orbital periods for CVs, $\text{P}_{\rm orb, CVs} \sim 80-360$\,min (see e.g. \citealt{Period_Range}) and the longest observed orbital period for a polar (V1309\,Ori with $478.96$\,min; \citealt{4}).

In these cases, the detection failed in finding the correct periodic signal due to factors such as the predominance of instrumental artefacts or noise. For instances where a spurious detection was obtained initially, the period search methods were reapplied with the search range restricted to 30–1000 minutes. This range extends beyond the typical range of orbital periods for CVs to encompass both the typical orbital periods of CVs and possible outliers, while still allowing the recovery of signals that may have been missed in the initial attempt, particularly due to the presence of dominant low-frequency power from red noise.

Detections within the expected range of orbital periods in CVs, but not in accordance with the literature reference, were subjected to further study as follows. 

When a detected period value is in disagreement with other measurements, that is, detections made by other numerical methods or in other observed TESS light curves (if they exist), 
and those other measurements  align with the literature value,
an individual inspection of the  measurement was done to confirm whether it constituted a faulty detection, for example whether the measured period arose from noise. In practice, in all these cases, the deviating period was discarded. If no additional detection in agreement with the literature value was made from the TESS light curves and it is not clear if the detection was faulty, the deviating measurement was retained and a detailed analysis  was done on an individual basis in order to assess if the identified signal could correspond to the orbital period. These cases are discussed in Sect.~\ref{Asynchronous} and Sect.~\ref{Outliers}. 

Once the pathological cases were disregarded and the periods corresponding to $1/2 \,\text{P}_{\text{orb}}$ were identified and corrected (see Sect.~\ref{HalfPeriodsCorrection}), the final adopted orbital period values were calculated. 
For those systems which were repeatedly observed, that is, systems for which there exist several observed two-minute cadence TESS light curves,  final values of the orbital period were computed for each detection method by taking the average among all repeated observations,

\begin{equation}
 \text{P}_{\rm orb, k} := \dfrac{\sum_{j}^{n_{k}}\text{P}_{\rm orb; k, j}}{n_{\rm k}}.
\label{eq:X}
\end{equation}

\noindent Here $\text{P}_{\rm orb, k, j}$ is the orbital period for the method $k$ (accounting for Lomb-Scargle periodogram, the autocorrelation function, sine fitting, and Fourier power spectrum analysis) and the observed light curve $j$, $n_{k}$ is the total number of detections in different observed light curves with the detection method $k$, and $\text{P}_{\rm orb, k}$ is the orbital period associated with a specific detection method $k$. As mentioned above, a given detection method may fail to detect a period in some light curves, that is, $n_{k}\leq n_{obs}$, where $n_{obs}$ is the number of available two-minute cadence TESS light curves of the system. 

The uncertainties were calculated as the 1-$\sigma$ confidence interval of the different period measurements using the standard error scaled by the appropriate Student’s $t$-factor. 
This was done separately for each of the four implemented  period search methods, thus, providing an estimate of the uncertainty for each numerical method,

\begin{equation}
\Delta \text{P}_{orb, k} := t_{0.6827,\,n_k-1} \cdot \dfrac{\sigma\left( \text{P}_{orb; k, j} \right)}{\sqrt{n_k}},
\label{eq:uncertainty_method}
\end{equation}

\noindent where $\Delta \text{P}_{orb, k}$ is the orbital period uncertainty associated to a specific detection method $k$, $\sigma$ denotes the standard deviation, and $t_{0.6827,\,n_k-1}$ is the two-tailed critical value of the Student’s $t$-distribution for a confidence level of 68.27\% and $n_k-1$ degrees of freedom. If for a numerical method $k$, only one measurement exists, that is, $n_{k}=1$, the standard deviation is not defined, and therefore no uncertainty $\Delta \text{P}_{orb, k}$ was associated to that numerical method. 

After associating for a given target a unique orbital period and, if possible, an uncertainty to each detection method, the final adopted orbital period and its uncertainty are selected as the pair with the lowest relative uncertainty,

\begin{equation}
\left( \text{P}_{\text{orb, final}},\, \Delta \text{P}_{\text{orb, final}} \right) := \arg\min_{k} \left( \dfrac{\Delta \text{P}_{\text{orb}, k}}{\text{P}_{\text{orb}, k}} \right),
\label{eq:best_period}
\end{equation}

\noindent which corresponds to adopting the most precise estimate among the implemented methods.

For systems with only one available two-minute cadence TESS light curve, the final orbital period is adopted from the Lomb-Scargle periodogram. In these cases, the uncertainty is estimated using a Monte Carlo approach, as described in Sect.~\ref{sec.Uncertainties}. In the following, we discuss the $10$ systems where $\text{P}_{\rm orb,final}$ is inconsistent with the literature value.

\subsubsection{Asynchronous systems}
\label{Asynchronous}

In some polars, the spin period of the WD is not synchronised with the orbital period, leading to complex light curve variations. For some of these asynchronous systems, the spin period signal (hereafter denoted as the frequency $\omega=2\pi \text{P}_{\text{spin}}^{-1}$) can be more prominent than the orbital period (hereafter denoted as the frequency $\Omega=2\pi \text{P}_{\text{orb}}^{-1}$), and this may lead to wrong interpretations of the detected periodic signal. Additionally, the mismatch between the WD's spin and the orbital motion causes drifting of the accretion spots across the WD's surface and 'pole-switching' accretion, that is, the accretion channel ends alternately on the two poles of the WD. This is because in discless CVs the accretion stream feeds material along the WD’s magnetic field lines onto the closest magnetic pole. Since the secondary star and the stream co-rotate with $\Omega$, the WD's dipole field rotates at an apparent angular frequency $\omega-\Omega$, as seen in a frame co-rotating with the binary’s orbital motion at frequency $\Omega$. Consequently,  
as the phase difference between $\omega$ and $\Omega$ evolves in time, accretion alternates between the two magnetic poles. 
To a first approximation, the 'pole-switching' phenomenon occurs when the spin and orbital phase differ by 90º.  
This causes a brightness modulation at the beat period $2\pi\left(\omega-\Omega\right)^{-1}$, as each magnetic pole is actively accreting for half this beat period. Additional considerations about the system's geometry  explain the modulations at additional beat frequencies (e.g. $2\omega-3\Omega$, $2\omega-\Omega$, $4\omega-3\Omega$) between $\omega$ and $\Omega$, also denominated as side-band frequencies (see \citealt{Wynn&King}, \citealt{Wang_2020} and \citealt{Beat_Freq_Asynchronous_Polars} for a detailed description).

The sample studied in this work comprises four  systems that display such asynchronous rotation. For these cases, a deeper analysis of their Lomb-Scargle periodograms was performed to distinguish between the spin period and the orbital period. In Table~\ref{table:asynchronous}, the orbital and spin periods measured in these systems are summarised. Additionally, all detected periodicities, together with their interpretations and S/N$_{\text{PSD}}$ values (see Appendix~\ref{sec:SNR}), are reported in Tables~\ref{table:BYCam}-\ref{table:Paloma} in Appendix~\ref{sec:periods_asynchronous}.
As discussed in Sect.~\ref{Reliability_Periods}, period detections with S/N$_{\text{PSD}}$ < 0.004 are considered highly unreliable. In the following, we describe these measurements in detail and we compare them to the results from previous studies of the same systems.

\begin{table}[h!]   
\centering
\caption{Orbital and spin periods of  asynchronous systems in the sample of polars observed with TESS in two-minute cadence. } 
\vspace{1mm}
\label{table:asynchronous}  
\resizebox{0.95\linewidth}{!}{
\begin{tabular}{c c c c}  
\hline 
  System  & Sectors & $\text{P}_{\text{orb}}$[min] & $\text{P}_{\text{spin}}$[min] \\ 
\hline
  BY Cam & 19, 59  & $201.220\pm 0.072$ & $199.08\pm 0.24$ \\
  IGR J19552+0044 & 54  & $83.0153\pm0.0086$ & $81.3011\pm0.0015$ \\
CD Ind & 1, 27  & $111.9719\pm0.0024$ & $110.9200\pm 0.0023$ \\
Paloma &  19 & $157.187\pm0.018$ & $136.235\pm0.025$ \\
\hline
\end{tabular}
}
\end{table}

\paragraph{BY\,Cam (TIC\,339374052)}
\label{sec:TIC_339374052}

Several periods were found for this system in previous observations (see e.g. \citealt{Mason_2} and references therein). The asynchronous rotation of the primary leads to brightness modulations at $\Omega$, $\omega$, and additional beat frequencies. The side-band frequency at $2\omega-\Omega$ is typically the most dominant signal and has been detected in multiple occasions for BY\,Cam (\citealt{Silber_1997}, \citealt{Mason_1998}, \citealt{Pavlenko}, \citealt{Honeycutt&Kafka}, and \citealt{Mason_2}).

\begin{figure*}
	\centering
	\includegraphics[width=1.0\linewidth]{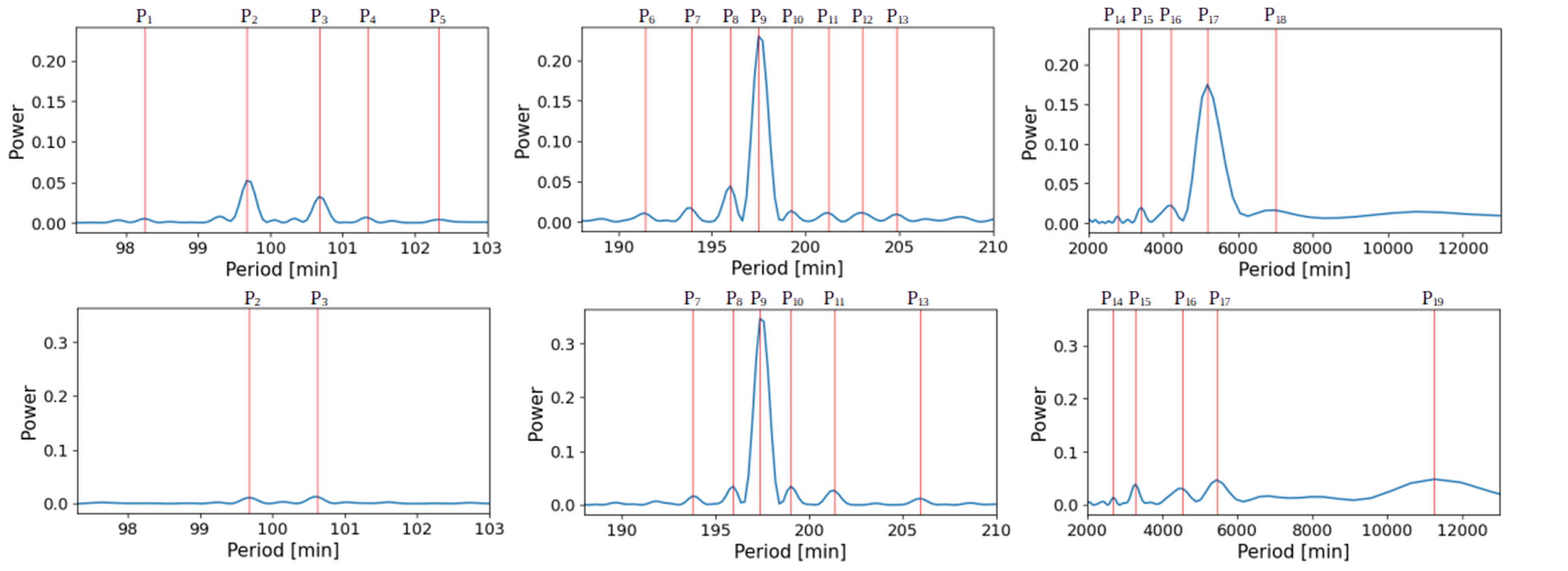}
	\caption{Zoomed-in view of the Lomb-Scargle periodograms of the TESS light curves of TIC\,339374052 (alias BY\,Cam). Top panels: Sector\,19. Bottom panels: Sector\,59. The red lines mark the positions of the different periodicities that were measured (see Table~\ref{table:BYCam}).}
	\label{Fig.TIC_339374052}
\end{figure*}

In our study, two TESS light curves were analysed, corresponding to sectors\,19 and 59 (see Fig.~\ref{Fig.TIC_339374052}).We detect signals corresponding to the orbital and spin periods, as well as the side-band period $2\pi \left(2\omega-\Omega\right)^{-1}$. Additionally, we identify multiple periodicities that we associate with side-band periods and their harmonics. Among all the detected signals, only the periods P$_{8}$, P$_{9}$, P$_{10}$, P$_{11}$, P$_{13}$, and P$_{18}$ have been previously reported (see \citealt{Mason_1998}, \citealt{Honeycutt&Kafka}, and \citealt{Mason_2})). Our results are presented in Table~\ref{table:BYCam}.  With the adopted interpretation of the different periods, we follow \cite{Mason_2}, in which the interpretation of the spin period is based on X-ray and polarisation analyses (see \citealt{Mason_1989} and \citealt{Mason_1998}) and the interpretation of the orbital period is based on spectroscopic studies (see \citealt{Mason_1996} and \citealt{Schwarz_2005}). 

\paragraph{IGR\,J19552+0044 (TIC\,228975750)}
\label{sec:TIC_228975750}

\cite{83} determined a  period $\text{P}_{\text{spin}} = 81.3$\,min using ground-based multi-band timeseries photometry and a period
$\text{P}_{\text{orb}} = 83.6$\,min from ground-based optical spectroscopy. 
 
\begin{figure}
	\centering
	\includegraphics[width=0.85\linewidth]{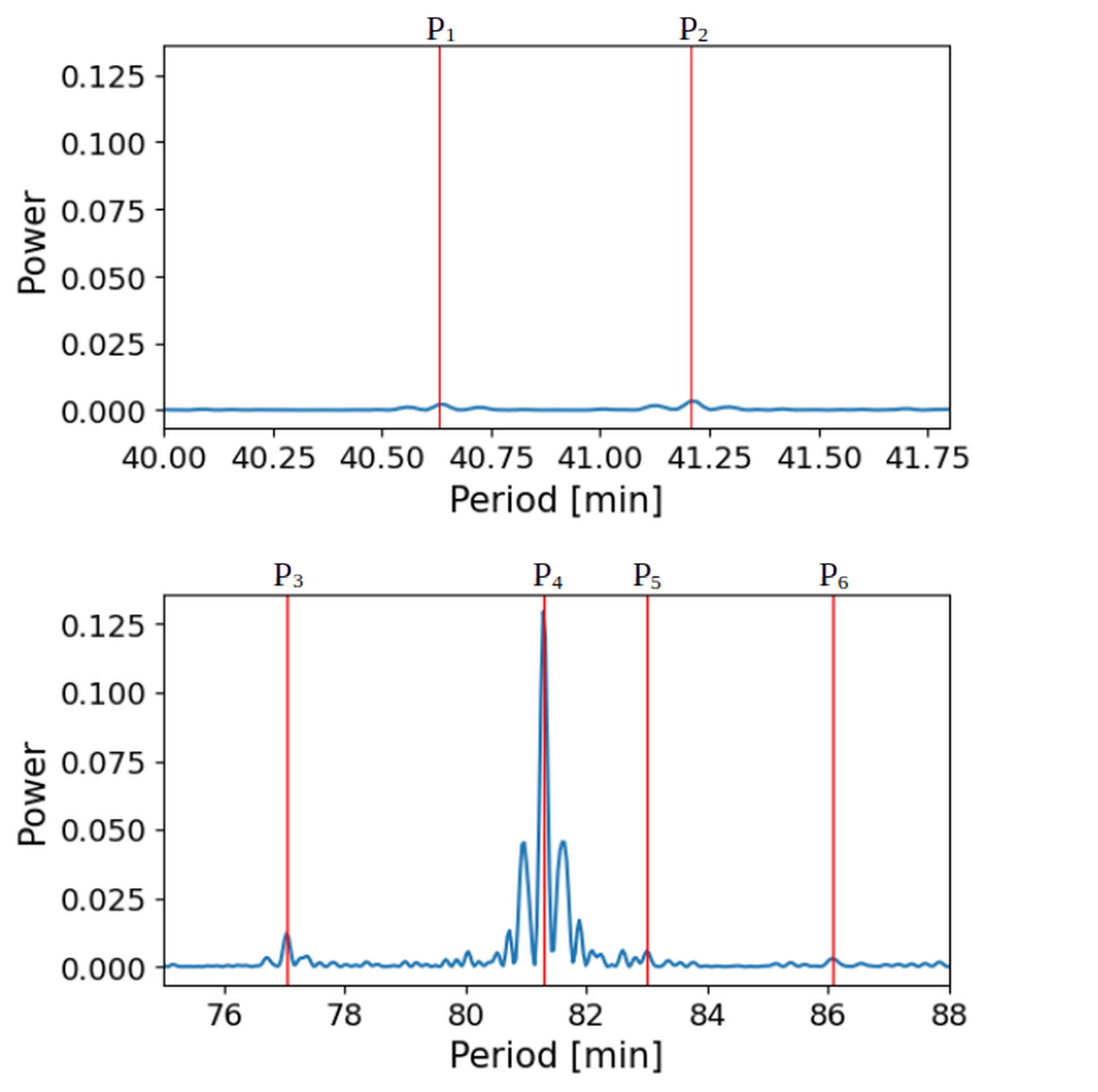}
	\caption{Zoomed-in view of the Lomb-Scargle periodogram of the TESS light curve of TIC\,228975750 (alias IGR\,J19552+0044) (sector\,54). The red lines mark the positions of the different periods that were measured (see Table~\ref{table:IGRJ19552}).}
	\label{Fig.TIC_228975750}
\end{figure}

In our work, the sector\,54 TESS light curve of IGR\,J19552+0044 revealed a distinct frequency peak at $81.3011\pm0.0015$\,min and a smaller signal at $83.0153\pm0.0086$\,min, corresponding to the periods $\text{P}_{4}$ and $\text{P}_{5}$ in the bottom panel of Fig.~\ref{Fig.TIC_228975750}, respectively. Consistent with the findings of \cite{83}, we identify the first signal with the spin period, $\text{P}_{\text{spin}}=81.3011\pm0.0015$\,min,  whereas the second is tentatively ascribed to the orbital period, $\text{P}_{\text{orb}}=83.0153\pm0.0086$\,min. However, given the low significance of this frequency peak, its presence is questionable.

We find four additional low-power signals (see Fig.~\ref{Fig.TIC_228975750} and Table~\ref{table:IGRJ19552}) that have not been reported in the previous literature.   
We interpret the periodicites $\text{P}_{1}$ and $\text{P}_{2}$ as
half the spin period and half the orbital period, respectively. \cite{83} found that the amplitude of the photometric variability was wavelength-dependent. Along with the changing shape of the system’s spectral continuum, this was regarded as evidence of cyclotron radiation. Following this  
interpretation,  
we assume that the variability at the spin period originates from cyclotron radiation emitted by the magnetic white dwarf, naturally explaining the signal at $1/2\,\text{P}_{\text{spin}}$ (see Sect.~\ref{subsect:half_period_mechanism}). 
The signal at half the orbital period is likely due to ellipsoidal modulation. The other two periods, $\text{P}_{3}$ and $\text{P}_{6}$, are interpreted as the side-band periods $2\pi \left(4\omega-3\Omega\right)^{-1}$ and $2\pi \left(3\Omega-2\omega\right)^{-1}$, respectively (see Table~\ref{table:IGRJ19552} in Appendix~\ref{sec:periods_asynchronous}).

\paragraph{CD\,Ind (TIC\,231666244)}
\label{sec:TIC_231666244}

\cite{49} found three significant periods in the TESS sector\,1 light curve: a period $\text{P}_{\text{orb}}=112$\,min interpreted as the orbital period, a period $\text{P}_{\text{spin}}=110.8$\,min interpreted as the spin period, and a period $2\pi\left(2\omega-\Omega\right)^{-1}=109.6$ min interpreted as a side-band period. Additionally, their analysis revealed multiple side-band periods and its harmonics in the power spectrum.

\begin{figure*}
	\centering
	\includegraphics[width=1.0\linewidth]{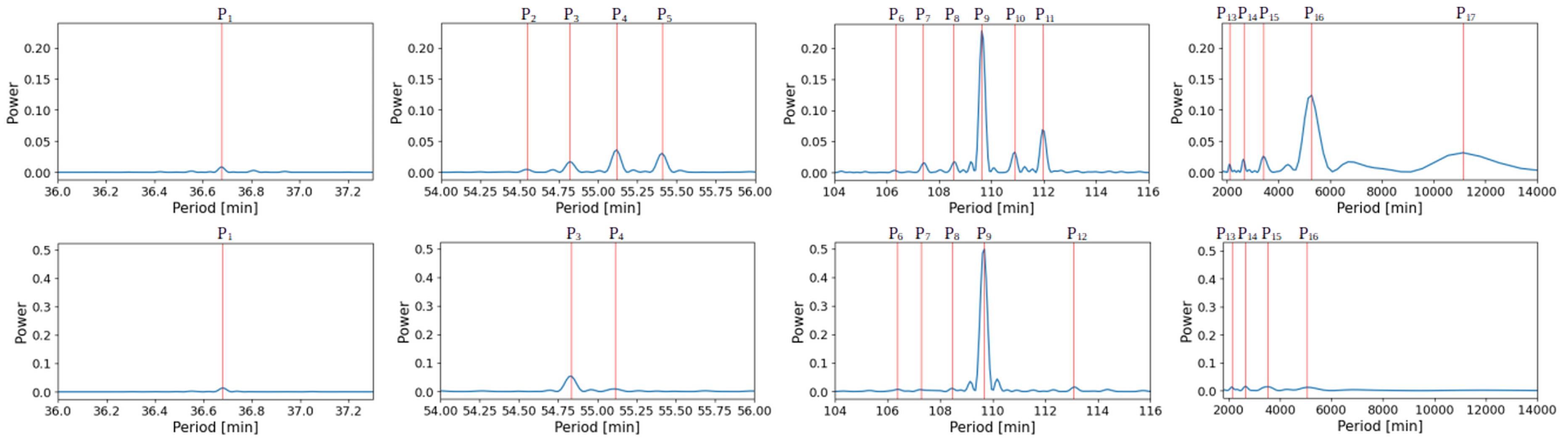}
	\caption{Zoomed-in view of  the Lomb-Scargle periodograms of the TESS light curves of TIC\,231666244 (alias CD\,Ind). Top panels: Sector\,1. Bottom panels: Sector\,27. The red lines mark the positions of the different periods that were measured (see Table~\ref{table:CDInd}).}
	\label{Fig.CD_Ind}
\end{figure*}

In this work, we analysed two TESS light curves corresponding to sectors\,1 and 27. We confirm the periods found by \cite{49} in sector\,1 and present the periods identified in sector\,27 (see Fig.\,\ref{Fig.CD_Ind} and Table~\ref{table:CDInd}). The adopted interpretation of the orbital and spin periods follows \cite{49}. Many of the periods observed in sector\,1, are also identified in sector\,27. The detection of these periods in   independent light curves further supports its physical origin and allows a more precise determination of their values. In sector\,27, the side-band period $2\pi \left(2\omega-\Omega\right)^{-1}$ (labelled $\text{P}_{\rm 9}$) appears with  higher significance, while other side-band periods are less prominent. Notably, the periods P$_{2}$, P$_{5}$, P$_{10}$, P$_{11}$, and P$_{17}$ are not present in sector\,27 (see Fig.\,\ref{Fig.CD_Ind}). The absence of P$_{10}$ and P$_{11}$ is particularly relevant, as they correspond to the spin and orbital periods, respectively. In addition, the side-band P$_{12}$, which is not clearly present in sector\,1, appears with greater significance in sector\,27. 

\paragraph{Paloma (TIC\,369210348)}
\label{sec:TIC_369210348}

\cite{Schwarz_1} proposed two different period identifications which were later verified by \cite{Littlefield_2} using TESS data. According to the interpretation given by these latter authors, the spin period of the white dwarf is $136.2$\,min and the orbital period is $157.2$\,min. 

\begin{figure}
	\centering
	\includegraphics[width=0.85\linewidth]{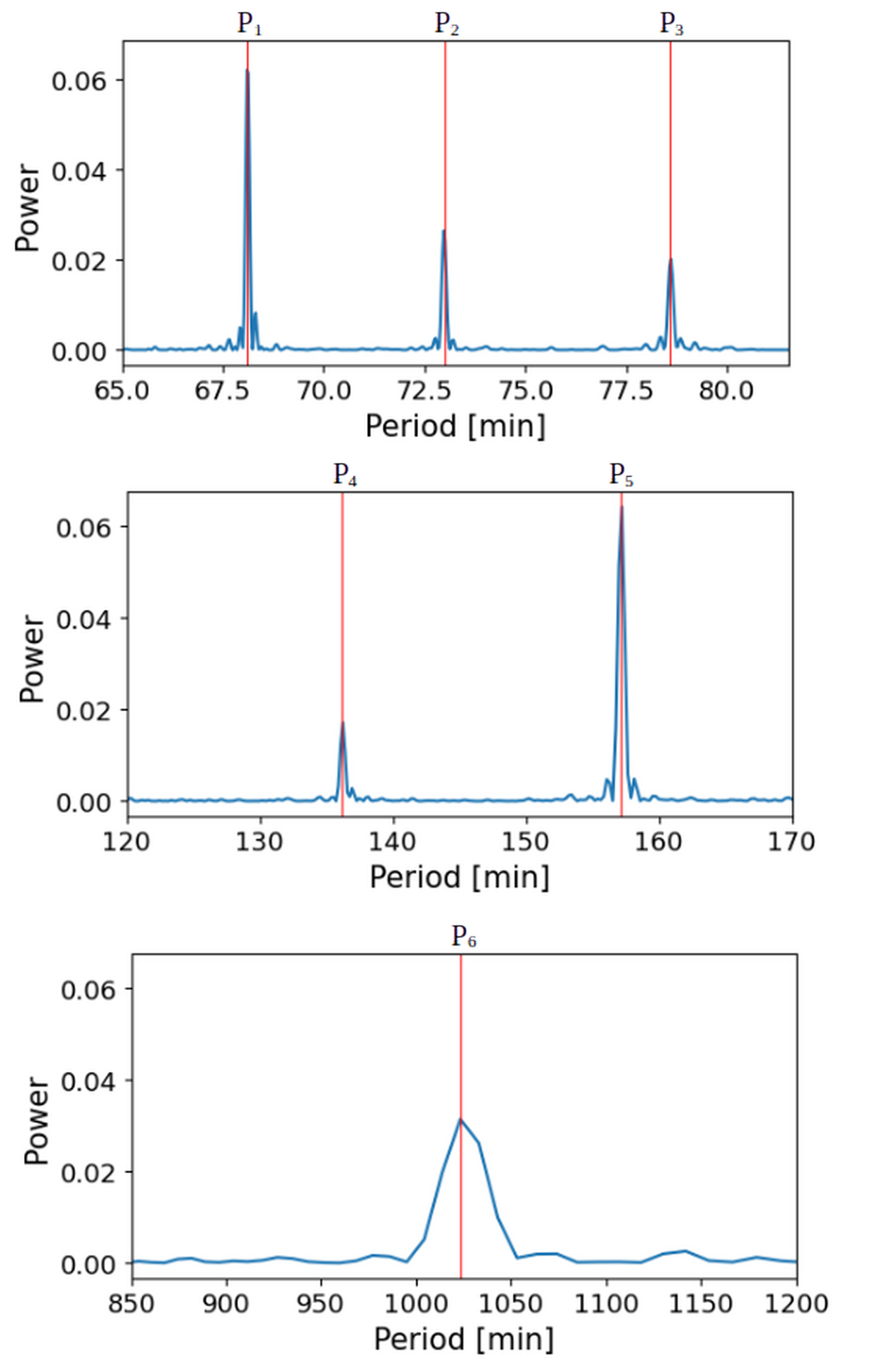}
	\caption{Zoomed-in view of the Lomb-Scargle periodogram of the TESS light curve of TIC\,369210348 (alias Paloma) (sector\,19). The red lines mark the positions of the different periods that were measured (see Table~\ref{table:Paloma}).} 
 	\label{Fig.TIC_369210348}
\end{figure}

We re-analysed the same TESS sector\,19 light curve and confirmed the periods found by \cite{Littlefield_2}. Here, we present the most significant periods (see Fig.\,\ref{Fig.TIC_369210348} and Table~\ref{table:Paloma}). 

\subsubsection{Outliers}
\label{Outliers}

Among the $95$ polars  analysed, for four systems the orbital periods  
detected with TESS exhibit values that  deviate significantly from values reported in the literature, as determined by their period differences exceeding the 95th percentile threshold (see Sect.~\ref{subsect:periods_results}). Here each of these outliers is examined individually. 

\paragraph{2XMM\,J154305.5-522709 (TIC\,1140118632)}
\label{sec:TIC_1140118632}

\noindent 
\cite{205} reported two periods based on {\it Chandra} X-ray data: P$_{1}=73.2\pm 4.83$\,min and P$_{2}=146.2\pm 15.95$\,min. \cite{31} reported a period of 143.4\,min based on multi-epoch $K$-band photometry, which appears consistent with the longer of the X-ray periods in \cite{205}. However, given the highly incomplete sampling of the infrared (IR) light curve, the reliability of this period in the IR data is questionable.

\begin{figure}
	\centering
	\includegraphics[width=0.8\linewidth]{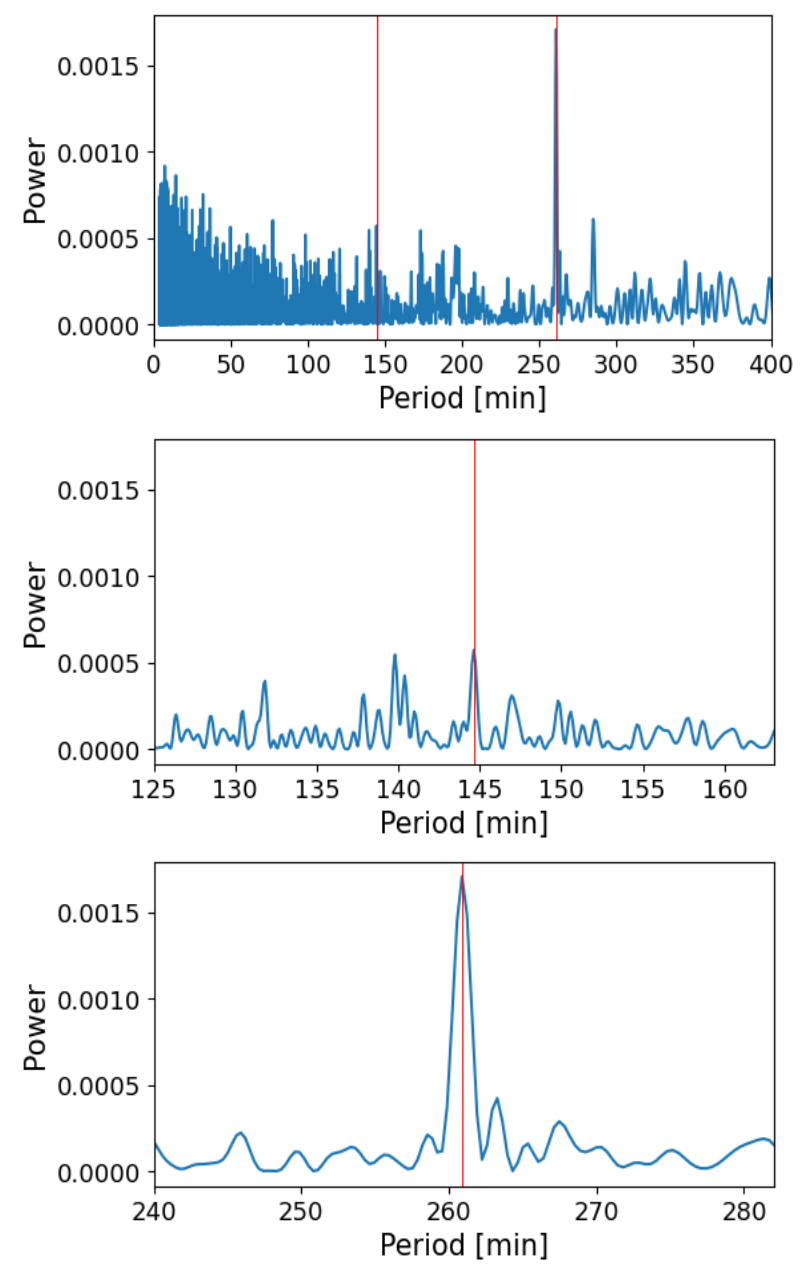}
	\caption{Top panel: Lomb-Scargle periodogram of the two-minute cadence TESS light curve of TIC\,1140118632 (alias 2XMM\,J154305.5-522709) (sector\,39). Centre panel: Zoomed-in view of the frequency peak located at 144.6\,min. Bottom panel: Zoomed-in view of the frequency peak located at $260.90\pm0.29$\,min. The red lines mark the positions of the frequency peaks.}
	\label{Fig.TIC_1140118632}
\end{figure}

In this work, a frequency peak corresponding to a period of $144.6$\,min is observed in the Lomb-Scargle periodogram of the only available two-minute cadence TESS light curve (sector\,39), but its distinction from noise is doubtful (see Fig.~\ref{Fig.TIC_1140118632}). A dominant frequency peak is 
located at $260.90\pm0.29$\,min (see Fig.~\ref{Fig.TIC_1140118632}). Notably, the same periodicity is detected by all four implemented period search methods. It is possible that this  modulation was not found by \cite{205} because their X-ray light curve spans only $\sim 500$\,min. 

This object is located in a crowded field along the Galactic plane. Therefore, we assessed the possibility of contamination by  neighbouring astrophysical objects contributing to the flux in the TESS light curve. Hereby, the methodology from \cite{Binks} was followed. Briefly, all \textit{Gaia} sources within a certain pixel radius of the target are considered and the contribution from these sources to the circular target aperture is given by (see \citealt{1965a_Biser}),
\begin{equation}
f_{\rm bg} = \sum_{i} 10^{-0.4 G_{i}}\, e^{-t_{i}} \sum_{n=0}^{n\to{\infty}}{\Bigg\{\frac{t_{i}^{n}}{n!}\bigg[1-e^{-s}\sum_{k=0}^{n}{\frac{s^{k}}{k!}}}\bigg]\Bigg\}\, ,
\label{eq:fbg}
\end{equation}
 
 \noindent where the outer summation includes all potential contaminant sources, $G$ represents the G-magnitude, $s=R^{2}/2\sigma^{2}$, $t = D^{2}/2\sigma^{2}$, $R$ is the radius of the aperture in pixels, $D$ is the distance between the potential contaminating object and the centre of the target aperture in pixels, and $\sigma$ is the full width at half-maximum of the TESS point spread function, which is roughly 0.65 pixels. 

The contaminating flux, $f_{\rm bg}$, is then compared with the flux from the target object within the aperture radius. The latter is approximately given by a Rayleigh distribution,
\begin{equation}
f_{\ast} = 10^{-0.4G}\left(1-e^{-s}\right).
\label{eq:Rayleigh}
\end{equation}

Considering an aperture radius of three pixels (63 arcsec), we find that the strongest contaminating source  contributes with a flux of $2.12\times10^{-5}$ and the total background contribution is $f_{\rm bg}\approx 2.62\times10^{-5}$. The target's flux is $f_{\ast}\approx 8.27\times10^{-8}$, yielding a contamination ratio of $f_{bg}/f_{\ast}\approx 3.17\times10^{2}$. With such a high contamination ratio, the possibility that the  periodic signal in the TESS light curve arises from contamination of a neighbouring source cannot be discarded. 
On the other hand, none of the known objects within a two-arcmin radius is of a type for which a period of $260$\,min can be expected.

Since the X-ray light curve presented by \cite{205} shows a clear periodic behaviour, the interpretation given here is that the correct orbital period is $\text{P}_{\text{orb}}=146.2$\,min from \cite{205}. In our analysis of the TESS light curve, we find a period of $260.90\pm0.29$\,min; however, due to the high contamination ratio it remains highly unreliable and such that follow-up studies are suggested.

\paragraph{GQ\,Mus (TIC\,908535248)}
\label{sec:TIC_908535248}

\begin{figure*}
	\centering
	\includegraphics[width=0.78\linewidth]{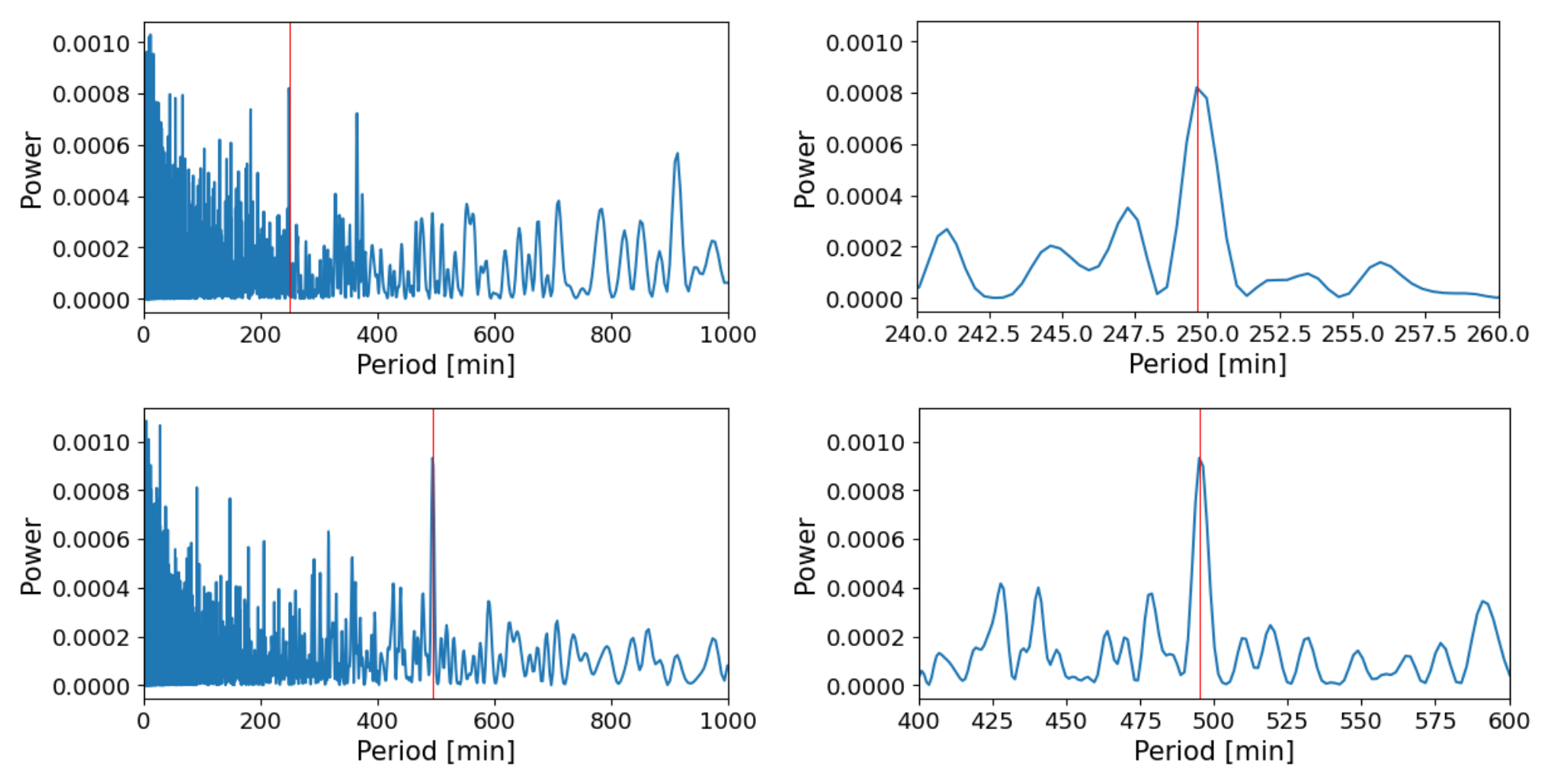}
	\caption{Left panels:  Lomb-Scargle periodograms of the TESS light curves  of TIC\,908535248 (alias GQ\,Mus) from  Sector\,37 (top) and Sector\,38 (bottom). Right panels: Zoomed-in view of  the region around the highest frequency peaks,  located at $249.64\pm0.57$\,min (top for Sector\,37) and at $495.1\pm2.3$\,min (bottom for Sector\,38). The red lines mark the positions of the frequency peaks.}
	\label{Fig.GQ_Mus}
\end{figure*}

\cite{24} reported a period of $85.5\pm 0.4$\,min based on data obtained with a photometer at the 1.6 m telescope of the Laboratório Nacional de Astrofísica at Brasópolis MG, Brazil. In our study, two TESS light curves were analysed, corresponding to sectors\,37 and 38. 
Different periods are found in the two light curves, namely
$\text{P}_{1}=249.64\pm0.57$\,min in sector\,37 (see top panels in Fig.~\ref{Fig.GQ_Mus}) and $\text{P}_{2}=495.1\pm2.3$\,min in sector\,38 (see bottom panels in Fig.~\ref{Fig.GQ_Mus}). No significant frequency peaks near $85$\,min were detected.
The two periods detected are roughly distinguished by a factor two (where $\text{P}_1$ is $\sim 4$\,min different to $1/2\,\text{P}_2$), suggesting that the periods could be physical.
However, both light curves present high levels of noise. Therefore, it cannot be excluded that these detections are spurious.
As a matter of fact, GQ Mus has a TESS magnitude of 18.796, and in Sect.~\ref{subsect:DetectionProbVSTmag} we argue that for TESS magnitudes $\geq$17 a substantial number of light curves are too noisy to reliably determine the orbital period. In fact, as shown in Fig.~\ref{Fig.SNR_vs_Magnitude} which is discussed in Sect.~\ref{subsect:DetectionProbVSTmag}, the periodic signals detected in the TESS light curves of GQ\,Mus present a S/N$_{\text{PSD}}$ (see Appendix~\ref{sec:SNR}) below the limiting value for a reliable orbital period detection. Thus, follow-up observations are required to confirm the orbital period for this system.

\paragraph{J0733+2619 (TIC\,94840820)}
\label{sec:TIC_94840820}

 \cite{63} identified  two potential orbital periods based on Catalina Real-Time Transient Survey (CRTS) optical light curves: $\text{P}_{1}=192.1$\,min and $\text{P}_{2}=200.9$\,min.  \cite{206} reported an orbital period of $200.29680\pm0.00029$\,min  based on photometric observations performed with the 1m Zeiss-1000 telescope of the Special Astrophysical Observatory of the Russian Academy of Sciences (SAO RAS), which aligns with the longer period reported by \cite{63}. 
 
\begin{figure}
	\centering
	\includegraphics[width=0.8\linewidth]{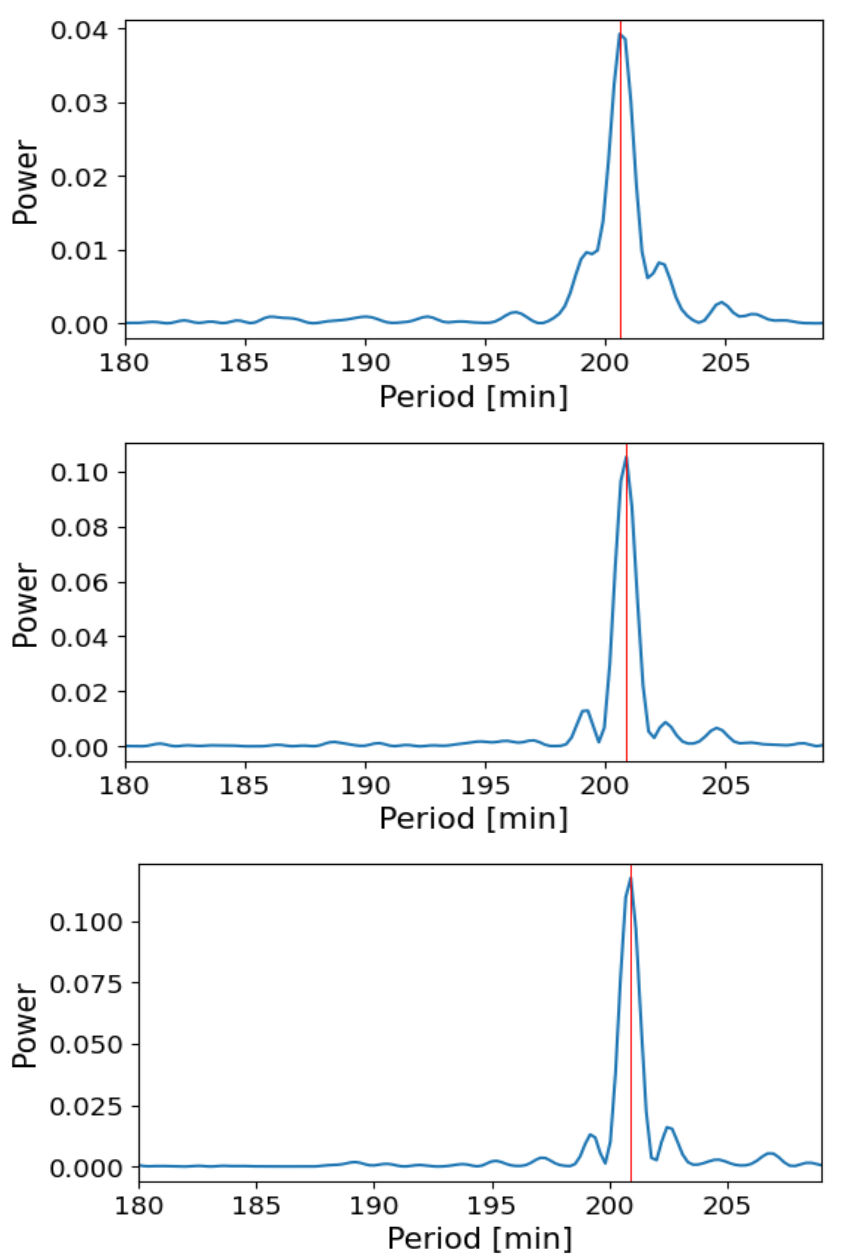}
	\caption{Zoomed-in views of the frequency peak at $\sim 200.8$\,min in the Lomb-Scargle periodograms of the TESS light curves of TIC\,94840820 (alias J0733+2619). Top panel: Sector\,44. Centre panel: Sector\,45. Bottom panel: Sector\,46. The red lines mark the positions of the frequency peaks.}
	\label{Fig.TIC_94840820}
\end{figure}

We studied three TESS light curves, corresponding to sectors\,44,\,45, and\,46. All three light curves present long term non-linear trends, likely driven by instrumental effects, that dominate the observed variability. For this reason, Lomb-Scargle periodograms were generated with a frequency range restricted to 0–1000 minutes. 
The $\sim 200.8$\,min period from \cite{63} is recovered with high significance in all three light curves (see Fig.~\ref{Fig.TIC_94840820}). Following the procedure described in Sect.~\ref{subsect:period_adopted}, our final adopted orbital period value is $\text{P}_{\text{orb}}=200.79\pm0.13$\,min. The period at 192\,min claimed by \cite{63} is not observed in the TESS light curves. Due to the much poorer data sampling of the CRTS visual light curves compared to the TESS light curves, our value for the (orbital)  
period measured on TESS data should be considered as more reliable; moreover, we are able to provide an uncertainty on this measurement based on multiple detections of the periodicity. The orbital period reported by \cite{206}, although not formally consistent with our result within the uncertainties, is in close agreement, further supporting the reliability of its detection.

\paragraph{J1424-0227 (TIC\,1002329407)}
\label{sec:TIC_1002329407}

\begin{figure}
	\centering
	\includegraphics[width=0.8\linewidth]{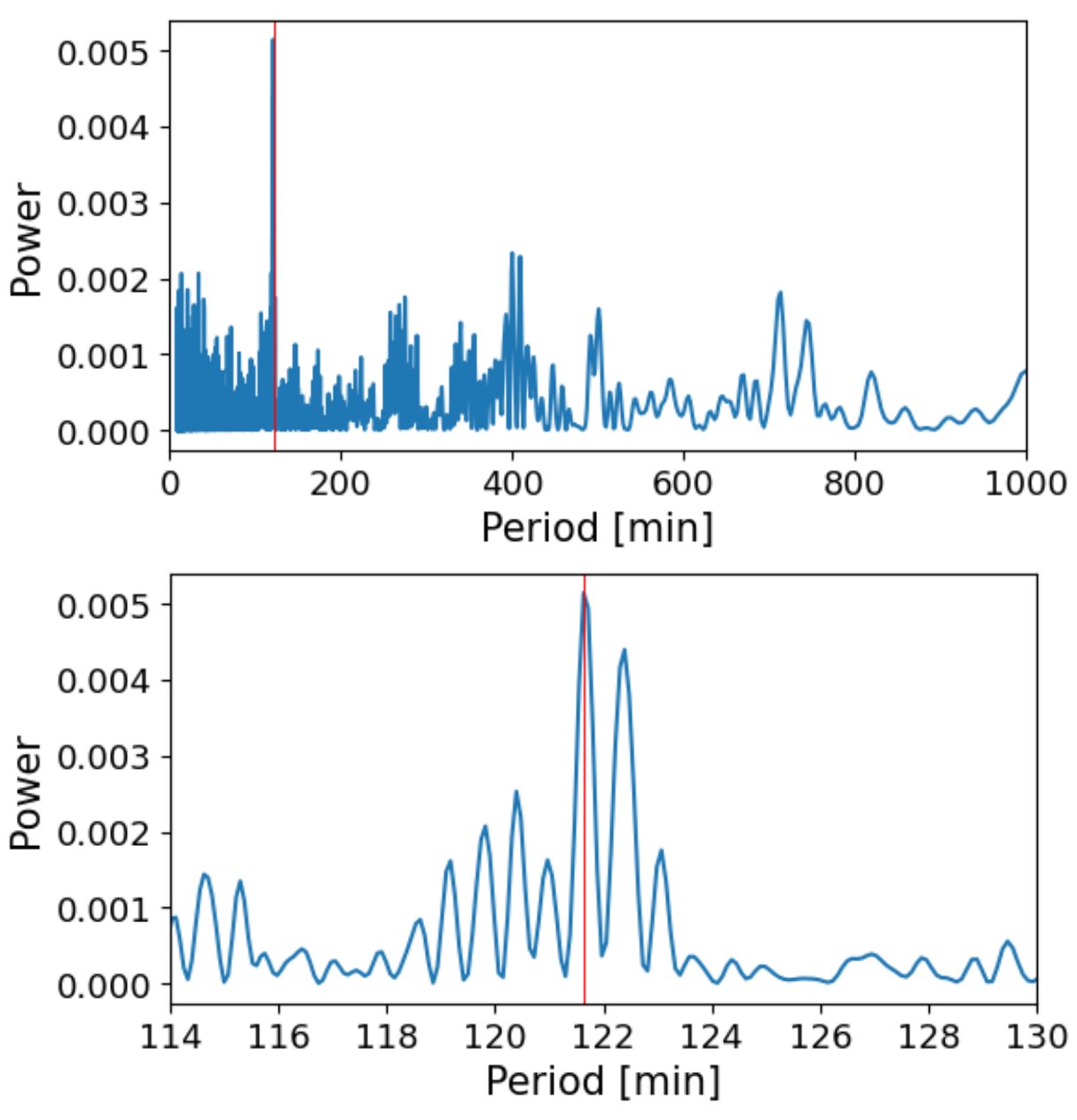}
	\caption{Top panel: Lomb-Scargle periodogram of the TESS light curve of TIC\,1002329407 (alias J1424-0227) (sector\,51). Bottom panel: Zoomed-in view of the frequency peak located at $121.636\pm0.063$\,min. The red lines mark the positions of the frequency peaks.}
	\label{Fig.TIC_1002329407}
\end{figure}

\cite{1} reported a period of 230.4\,min based on WISE data. Additionally, \cite{Marsh_outlier} reported a period of 223.92\,min using the ISIS spectrograph on the 4.2m William Herschel Telescope in the Canary Islands. Our analysis of the TESS light curve for J1424-0227 (sector\,51) revealed a period of $121.636\pm0.063$\,min (see Fig.~\ref{Fig.TIC_1002329407}). However, the  TESS light curve is noisy and it contains two data gaps in addition to the LAHO gap. 

The high noise level in the TESS light curve can be explained by the faintness of the system, which has a TESS magnitude of $17.541$, in the range where according to our analysis a period detection is potentially unreliable (see Sect.~\ref{subsect:DetectionProbVSTmag}). 
Under these circumstances, it becomes difficult to assess if the signal at $121$\,min has a physical origin. Considering that signals at $\sim 230$\,min or $\sim 223$\,min are not present in the TESS light curve, we suggest follow-up observations to confirm the orbital period. 

\subsubsection{Unsuccessful period measurements}
\label{Noisy_Cases}

\begin{figure}
    \centering	
    \includegraphics[width=0.8\linewidth]{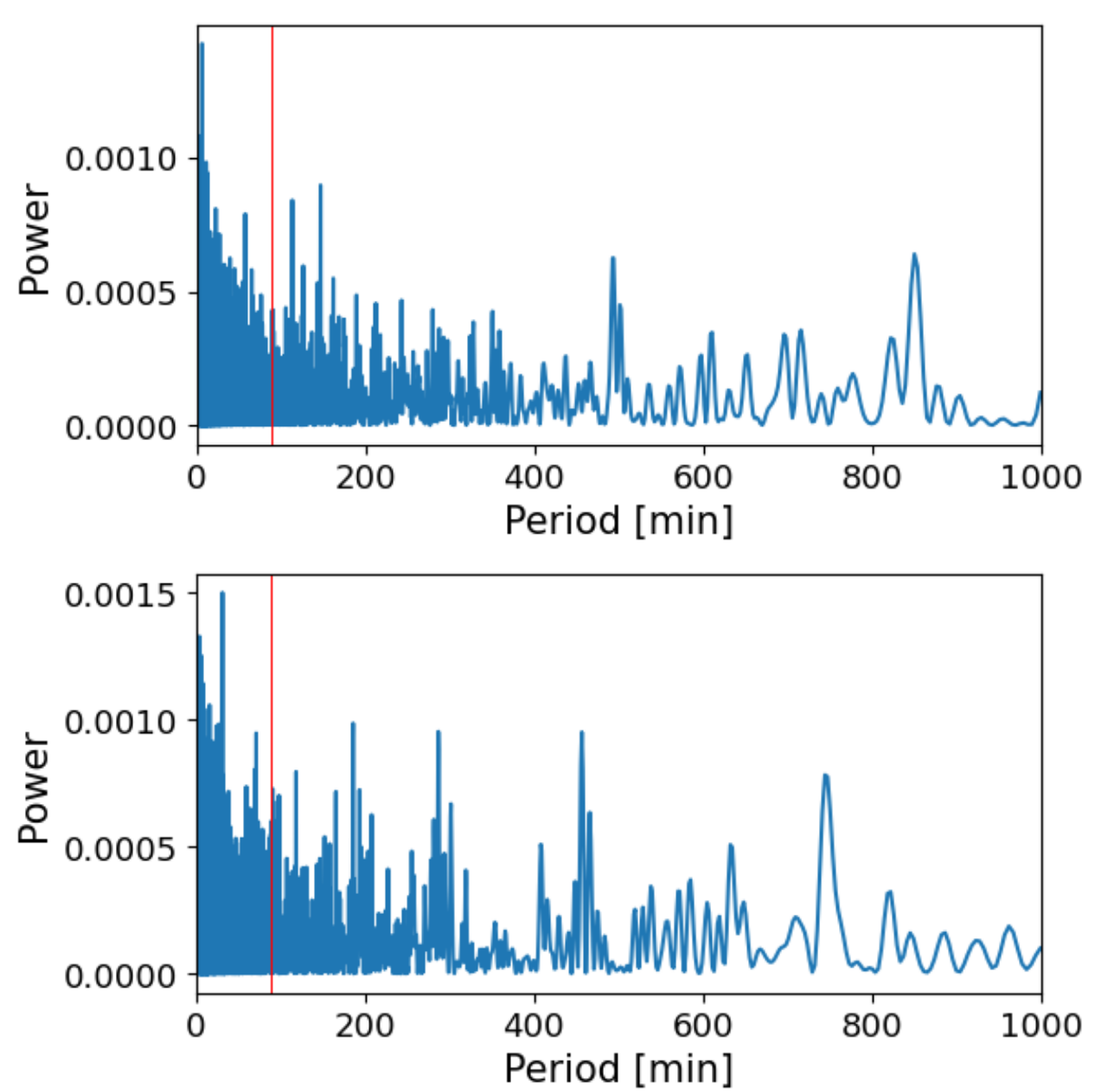}
    \caption{Top panel: Lomb-Scargle periodogram for TIC\,641322558 (sector\,58). Bottom panel: Lomb-Scargle periodogram for TIC\,953321496 (sector\,46). The low S/N impedes the identification of periodic signals. The red lines mark the positions of the orbital periods reported in the literature (see \citealt{64} and \citealt{25}).}
    \label{Fig.TIC_641322558_and_TIC_953321496}
\end{figure}

For two systems, J0257+3337 (TIC\,641322558) and V379\,Vir (TIC\,953321496), the S/N of their two-minute cadence TESS light curves did not allow us to recover the  (orbital) periods previously reported by \cite{64} and \cite{25}. Fig.~\ref{Fig.TIC_641322558_and_TIC_953321496} shows their corresponding Lomb-Scargle periodograms. 
 It is important to note that J0257+3337 and V379\,Vir have TESS magnitudes of 18.18 and 17.99, respectively. This is in accordance with our results presented in Sect.~\ref{subsect:DetectionProbVSTmag}, where we establish that for TESS magnitudes $\geq17$\,mag, period detections are highly unreliable. 

\section{Flattened  light curves}
\label{sec.Flat_LCs}

To determine the correlation between the noise levels in the studied light curves and the TESS magnitude, the light curves were 'flattened' removing  their periodic  modulations, thus leaving only the noise as the content of the light curve. 

The flattening methodology consists in generating smoothed versions of the observed light curves using the centred moving average algorithm (\citealt{Moving_Averages}). To conduct this analysis, an interactive code was developed, which allowed the precise adjustment of the smoothing window size for each individual light curve, and therefore ensured an optimal smoothing. Excessive smoothing would impede a correct estimation of the astrophysical periodic modulation, causing significant details to be lost. Conversely, insufficient smoothing would additionally capture noise and fail to adequately filter out random fluctuations. The appropriate window size for each light curve was determined empirically through visual evaluation of the resulting smoothed light curves for different smoothing window values. In future works, this process could be automatised by selecting the smoothing window size that produces residuals with respect to the original light curve that best approximate a normal distribution.

By subtracting the smoothed light curves from the original observed light curves, one obtains the flattened light curves, 

\begin{equation}
y_{\text{flat}}(t) = y(t)-y_{\text{smoothed}}(t).
\end{equation}

\noindent Finally, once the flattened light curves are obtained, the standard deviation, $S_{\rm flat}$, is computed. The standard deviation of flattened light curves, that is, light curves of which the (known) astrophysical signal(s) have been removed, serves as an estimation of the dispersion present in the noise. In \cite{Sflat_1}, we have introduced  $S_{\rm flat}$ to characterise the noise in photometric light curves of M dwarf stars. Here the astrophysical signal that was removed to obtain flat light curves are rotational modulations due to star spots and flares. We note that for the calculation of the parameter $S_{\rm flat}$ in the M dwarf light curves, another automated code was used (see \citealt{Raetz_2020}).

\begin{figure}
	\centering
	\includegraphics[width=1.0\linewidth]{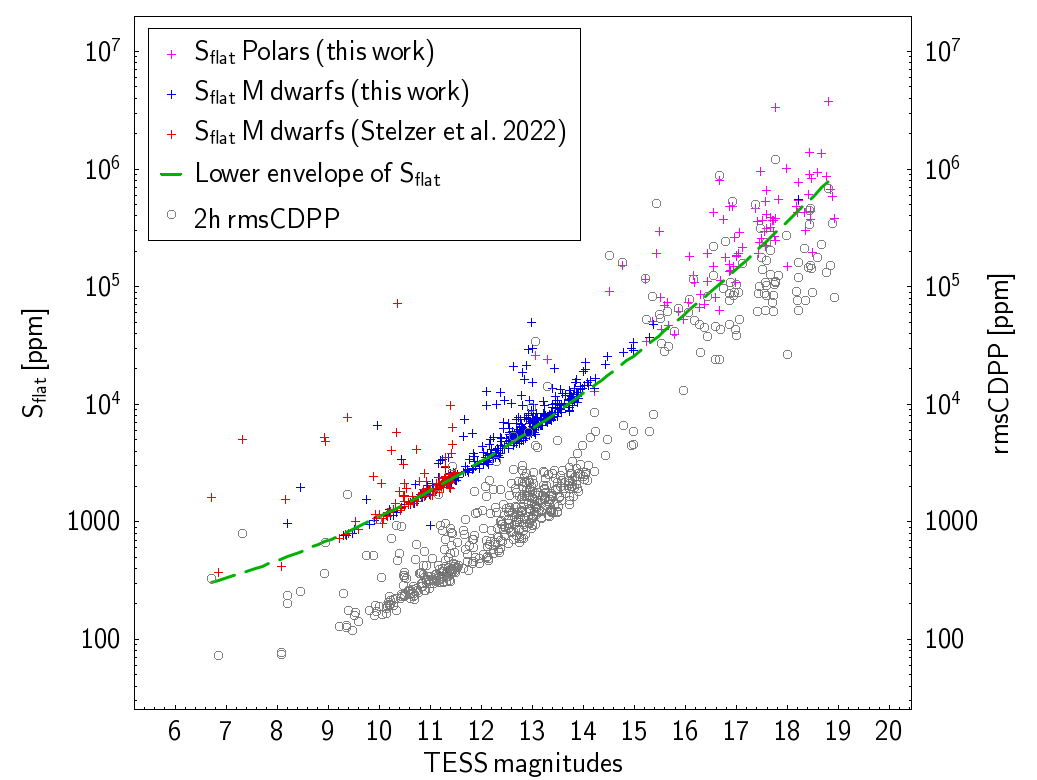}
	\caption{Relation between $S_{\rm flat}$ and TESS magnitude for polars (pink) compared to the same relation for  the M dwarf sample from  \cite{Stelzer_1} (red) and 
 a subsample of M dwarfs from \cite{Wilhelmina} (blue). The TESS light curves of the latter  were analysed by us (see Sect.~\ref{sec.Flat_LCs}). The lower envelope of the $S_{\rm flat}$ distribution (green dashed line) was  obtained by fitting a second-order polynomial to the lower $25$\,\% quantile of the data. The two-hour rmsCDPP values for all objects in the three samples are presented for comparison (see Sect.~\ref{sec.Flat_LCs} for details).}
	\label{Fig.Sflat_vs_Tmag}
\end{figure}

In our previous work on nearby M dwarfs observed in the K2 mission, we observed a correlation between $S_{\rm flat}$ and the {\it Kepler} magnitude $K_{\rm p}$ of the stars  (see \citealt{Sflat_1} and \citealt{Sflat_2}). An analogous study on TESS data was presented by \cite{Stelzer_1}. In Fig.~\ref{Fig.Sflat_vs_Tmag}, we show the relation between $S_{\rm flat}$ and the TESS magnitude for our sample of polars (black) together with the results from \cite{Stelzer_1} (red). For both samples the value of $S_{\rm flat}$ displayed in the figure is the mean of the values measured on the different TESS light curves of a given star.  
Since these nearby M dwarfs are much brighter than the known polars we closed the gap between the two samples with $340$ X-ray selected M dwarfs in the Southern Continuous Viewing Zone of TESS from \cite{Wilhelmina} (blue). For this subsample, only a single two-minute cadence TESS light curve was analysed per object.

A clear relation  over almost ten orders of magnitude with a well defined lower envelope is observed between the noise level of the flattened light curve and the brightness of the star. In \cite{Sflat_1} and \cite{Sflat_2}, it was observed that fast rotating M dwarfs (rotation periods $\text{P}_{\text{rot}}<10$ days) exhibit higher S$_{\rm flat}$ values than slow rotating stars (rotation periods $\text{P}_{\text{rot}}>10$ days). We attributed this to an additional underlying noise of astrophysical origin, such as unresolved mini-flares or mini-spots, for the fast rotators.  Similarly, the light curves of polars are affected by flickering, a stochastic variability occurring on timescales from milliseconds to hours, which constitute an intrinsic source of astrophysical noise in polars (see e.g. \citealt{Bruch_2021}, \citealt{flickering} and \citealt{Scaringi_2014}). While the periodic modulation due to the hotspot is effectively subtracted in our flattening process, the flickering component remains, contributing to varying noise levels and likely accounting for the higher dispersion observed in the sample of polars in Fig.~\ref{Fig.Sflat_vs_Tmag}.

Our $S_{\rm flat}$ parameter shares the rough profile of its dependence on the $T$ magnitude with the root mean square Combined Differential Photometric Precision (rmsCDPP) (\citealt{CDPP_1}, \citealt{CDPP_2}). Originally developed for the {\it Kepler} mission and later applied to TESS, the rmsCDPP parameter is optimised for the evaluation of the detectability of transit-like signals. In Fig.~\ref{Fig.Sflat_vs_Tmag}, we  also include for all objects in the three samples the 2-hour rmsCDPP measurements that are provided as a  pipeline product with the TESS light curve. For a given star, the values of rmsCDPP displayed in the figure represent the mean of the individual rmsCDPP measurements from the different TESS light curves.

An important difference  between the rmsCDPP and $S_{\rm flat}$ is that the former is calculated as the root mean square of the residuals relative to an average flux within short time intervals (in this case two hours).
This means that, for longer timescales, rmsCDPP averages out high-frequency noise. 
Moreover, in the calculation of rmsCDPP, the data is filtered to produce white noise, effectively whitening the time-correlated (red) noise. Thus, rmsCDPP measures the white noise in a light curve over a specific timescale, making it more suitable for evaluating the noise level in time-localised events such as transits. In contrast, $S_{\rm flat}$ captures the instantaneous scatter at each datapoint after removing the astrophysical periodic modulation(s), preserving both high-frequency and red noise in the flattened light curves. Therefore, it is better suited for assessing the overall noise level affecting period analysis. The  contribution of high-frequency noise to the scatter measured by $S_{\rm flat}$, which is averaged out in the calculation of rmsCDPP,  explains the systematically higher values compared to rmsCDPP. This is consistent with the idea that our flattened light curves still comprise astrophysical noise sources, possibly manifest in the form of high-frequency and red noise.

To quantify the lower envelope in Fig.~\ref{Fig.Sflat_vs_Tmag}, we divided the data into eight bins based on TESS magnitudes. Within each bin, we selected the $S_{\rm flat}$ values below the 25th percentile to represent the lower envelope of the distribution. These selected data points were then used to perform a second-order polynomial fit:
\begin{equation}
S_{\rm flat, env}(T)=aT^{2}+bT+c.
\end{equation}
Here $a$, $b$, and $c$ are the fitted coefficients. The fitting was performed using the Levenberg-Marquardt algorithm and to estimate the uncertainties in the fitted parameters, a bootstrapping approach was followed. That is, the data was resampled with replacement obtaining a total 10000 pseudo-samples, and the fitting procedure was repeated for each pseudo-sample to obtain a distribution for each coefficient. The standard deviation of these distributions was taken as the uncertainty in the corresponding parameter. The best-fit coefficients and their uncertainties are 
$a=0.01290\pm0.0021$,  
$b=-0.047\pm0.056$, and  
$c=2.22\pm0.38$.  

\section{Reliability of the detected periods}
\label{Reliability_Periods}

The detectability of orbital periods is significantly influenced by the S/N of the observed light curves. In this section, we describe  the methodology and the results of an investigation aimed at determining the limiting S/N necessary for a reliable orbital period detection in TESS light curves of CVs. As specified in Appendix~\ref{sec:SNR}, in this analysis, the S/N refers to the signiﬁcance of that frequency peak in the power spectral density (PSD) of a light curve that is associated to the orbital period. 
In the following we denote this signal-to-noise ratio as
S/N$_{\text{PSD}}$.

\subsection{Simulation-Recovery Test}
\label{subsect:Simulation-Recovery_Test}

To assess the limiting S/N$_{\text{PSD}}$, simulations based on an observed TESS light curve and resembling its statistical characteristics, namely the PSD and the probability density function (PDF),  were performed (see Fig.~\ref{Fig.Simulated_Light_Curve} in Appendix~\ref{sec:simulation}). The chosen light curve is the observation of TIC\,840418301 (alias IL\,Leo) in sector\,48. This light curve presents only one frequency peak in its PSD (see Fig.~\ref{Fig.SNR_psd} in Appendix~\ref{sec:SNR}), which greatly simplifies the calculation of S/N$_{\text{PSD}}$, making it appropriate as a basis for the simulations. 
As discussed in Appendix~\ref{sec:Normality}, this light curve has normally distributed noise, which is a requirement for the application of our simulation method (see Appendix~\ref{sec:simulation}). 
This observed light curve was selected as the basis for simulations and served as the 'true' reference signal with a $\text{P}_{\text{orb}}=82.4$\,min.

Synthetic replicates of the selected light curve were generated with varying levels of Gaussian noise. The S/N$_{\text{PSD}}$ was measured in each simulated light curve and noise was added to cover several S/N$_{\text{PSD}}$ intervals, ensuring a systematic evaluation of the detection performance.  
Specifically, seven S/N$_{\text{PSD}}$ intervals were defined and $50$ light curves were simulated for each S/N$_{\text{PSD}}$ interval. The size of the  S/N$_{\text{PSD}}$ intervals has been adjusted manually so as to provide higher resolution in the region of lower S/N$_{\text{PSD}}$ values. This provides a precise constraint of the limiting S/N$_{\text{PSD}}$ for orbital period detection. 
For each simulated light curve, the four period search methods were applied attempting to recover the orbital period of TIC\,840418301. The success rate of recovering the orbital period was recorded for each S/N$_{\text{PSD}}$ interval, and allowed us to estimate a probability of detection, which is defined as a frequentist probability, that is, the ratio of cases where the correct orbital period was recovered to the total number of attempts    within a given S/N$_{\text{PSD}}$ interval, which is $50$.

In a simulated light curve, the orbital period is reproduced stochastically, meaning that the periodic signal can deviate slightly from the original one. To evaluate if the orbital period was successfully recovered from a simulated light curve, the variability range with which the orbital period is reproduced in the simulated light curves was assessed. To this end, $50$ simulated light curves were generated without introducing extra white noise. 
After performing the period search on them, the standard deviation of the $50$ period values was calculated, yielding $\sigma = 0.22$\,min. 
Following this result,  a recovery attempt in a simulated light curve is considered as successful if the detected orbital period falls within the interval $\left(\text{P}_{\text{orb}}-3\sigma, \text{P}_{\text{orb}}+3\sigma\right)$, where $\sigma=0.22$\,min and $\text{P}_{\text{orb}}=82.4$\,min for TIC\,840418301 in sector\,48. 

\begin{figure}
	\centering
	\includegraphics[width=0.97\linewidth]{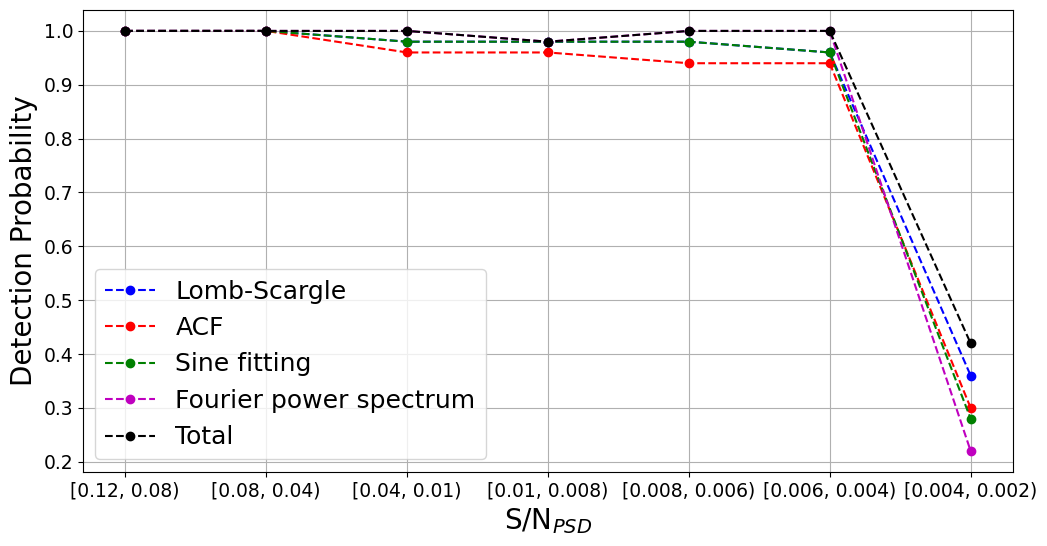}
	\caption{Detection probability as a function of the S/N$_{\text{PSD}}$ intervals.  In the interval notation used, a round bracket `~)~' indicates an open interval, excluding the ending point, while a square bracket `~[~' indicates a closed interval, including the starting point.}
\label{Fig.Prob_vs_SNR}
\end{figure}

The results of this simulation-recovery test are displayed in Fig.~\ref{Fig.Prob_vs_SNR} which shows the period detection probability for the seven S/N$_{\text{PSD}}$ ranges. Results are shown separately for each of the four detection methods as well as for the total effective detection probability which accounts for every successful recovery independent of the period search  method used.  

All methods maintain a high performance, with a detection probability above $90$\% in the recovery of the orbital period for six out of the seven S/N$_{\text{PSD}}$ intervals. The very high probability in the successful recovery of orbital periods across most of the S/N$_{\text{PSD}}$ range is followed by a steep decrease in the detection probability for the lowest S/N$_{\text{PSD}}$ interval [0.004, 0.002). In this S/N$_{\text{PSD}}$ range, the detection probability for all methods drops to below 40\%, and the total effective detection probability falls below 50\%. We conclude that this S/N$_{\text{PSD}}$ level renders period detections  highly unreliable, and additional observations might be necessary to confirm periods detected under such conditions. Consequently, the limiting ${\rm S/N_{PSD}}$ for a reliable orbital period detection in TESS light curves of CVs with the use of the numerical methods presented in this work corresponds to $({\rm S/N_{PSD})_{min}} =0.004$. 
Below we compare this threshold to the observed TESS data of the sample of polars. 

\subsection{Detection probability and TESS magnitude}
\label{subsect:DetectionProbVSTmag}

\begin{figure}
    \centering
    \includegraphics[width=0.95\linewidth]{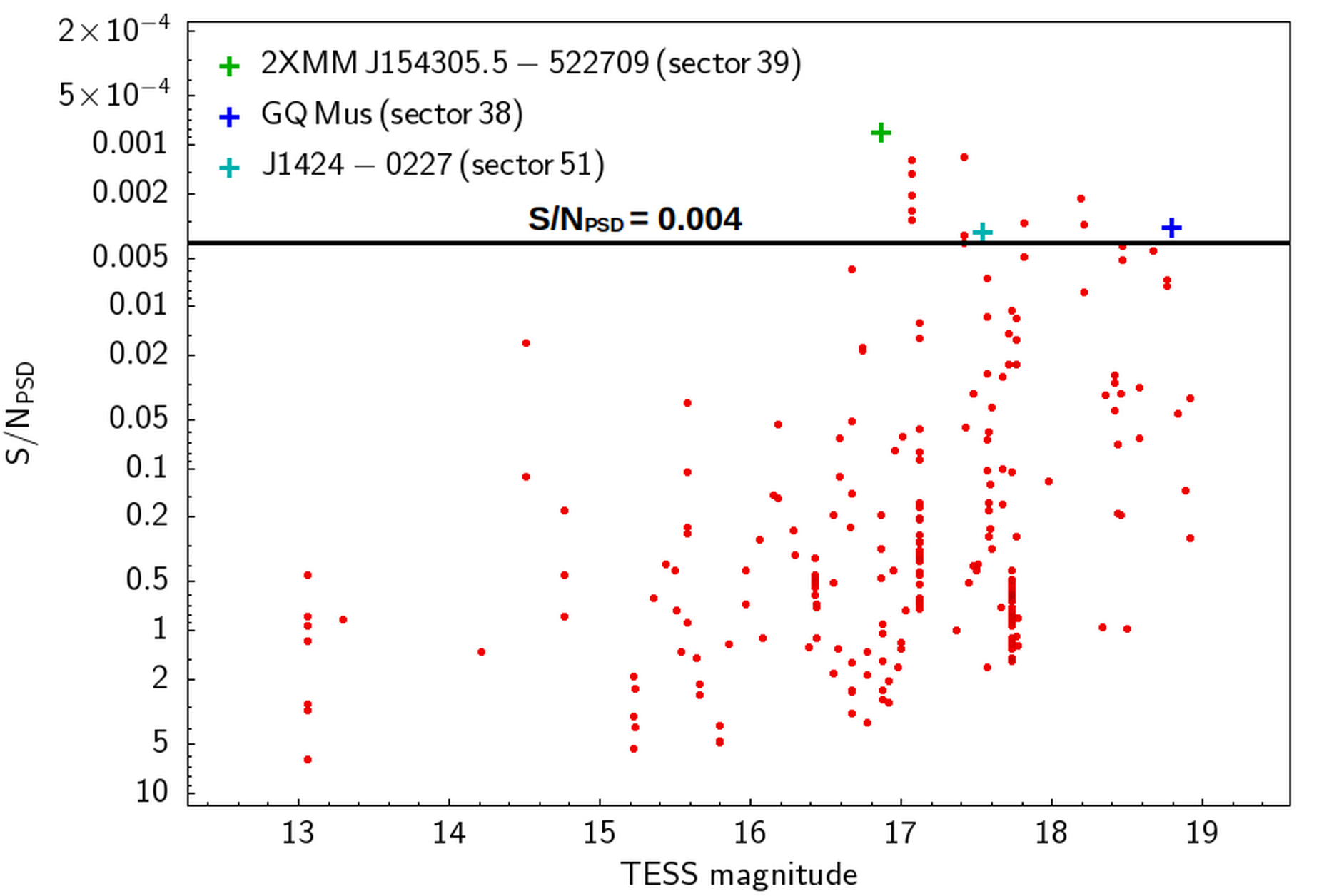}
    \caption{S/N$_{\text{PSD}}$ of observed TESS light curves as a function of the TESS magnitude. The black horizontal line indicates the limiting S/N$_{\text{PSD}}$ for a reliable orbital period detection,  S/N$_{\text{PSD}}$ = 0.004, determined with simulations (see Sect.~\ref{subsect:Simulation-Recovery_Test}). In Sect.~\ref{Outliers}, periods were provided for 2XMM J154305.5-52270,  GQ Mus, and J1424-0227 (marked here with different plotting symbols and colours), but our new periods as well as previous values presented for them in the literature remain questionable, consistent with the location of these systems beyond the ${\rm S/N_{\rm PSD}}$ threshold.} 
\label{Fig.SNR_vs_Magnitude}
\end{figure}

 The ${\rm S/N_{PSD}}$ value of observed TESS light curves is expected to decrease for fainter systems, that is systems with higher TESS magnitude. We have verified this on the sample of polars. The result is shown in Fig.~\ref{Fig.SNR_vs_Magnitude}. By comparing this with the ${\rm S/N_{\rm PSD}}$ threshold determined in Sect.~\ref{subsect:Simulation-Recovery_Test} we can determine a limiting TESS magnitude for orbital period detection in the sample.

The measurement of S/N$_{\text{PSD}}$ in the observed light curves was performed using the same methodology applied for the simulated light curves (see Appendix~\ref{sec:SNR}) but with some additional considerations.
The TESS light curve of TIC\,840418301 (sector\,48) which we used to produce simulated light curves presented only one periodic signal, which was ascribed to the orbital period. In our sample of polars, the situation is not that simple and other periodic signals—including $1/2\, \text{P}_{\rm orb}$ as discussed above—may be present in the PSD. Such additional signals must be removed. To this end we implemented the following criteria: 
\begin{enumerate}

\item If the frequency peak at $1/2 \,\text{P}_{\text{orb}}$ is present in the PSD, only the stronger of the $1/2 \,\text{P}_{\text{orb}}$ and the $\text{P}_{\text{orb}}$ frequency peak is associated with the signal, since this would be the signal that would be predominantly detected. In the noise-only signal construction, both frequency peaks are removed from the PSD.

\item In the estimation of the noise power, not only the frequency peak associated with the signal is removed from the PSD, but also additional short periodic signals ($P\sim 0-500$\,min) that may have their origin in additional astrophysical processes, for example the spin period in asynchronous systems, or arise from phenomena like beat frequencies or aliasing. The identification of such periodic signals was performed through visual inspection of each individual light curve. In practice, aliased frequency peaks are straightforward to identify as they appear as a series of harmonics, narrow and regularly spaced in the low-period region of the PSD. Additional signals, such as the beat frequencies normally appear as high-power and narrow peaks. These peaks would lead to an overestimation of the noise if not removed. Frequency peaks that were ambiguous and could plausibly originate from noise were retained.
\end{enumerate}

Figure~\ref{Fig.SNR_vs_Magnitude} presents the measured S/N$_{\text{PSD}}$ values from the observed TESS light curves against the TESS magnitudes of the corresponding objects. We note that the estimation of S/N$_{\text{PSD}}$ requires the presence of a periodic signal in the PSD. In some cases, a periodic signal appears in the Lomb-Scargle periodogram but is absent from the PSD calculated as described in Appendix~\ref{sec:SNR}.  This typically occurs in noisy light curves, suggesting that the frequency peak in the Lomb-Scargle periodogram may have arisen purely from noise. Applying the methodology of Appendix~\ref{sec:SNR}, a periodic signal is present in the PSD of 223 out of 235 TESS light curves, hence we have 223 S/N$_{\text{PSD}}$ values. The remaining cases do not appear in Fig.~\ref{Fig.SNR_vs_Magnitude}.
 
Figure~\ref{Fig.SNR_vs_Magnitude} confirms the expected trend: as the TESS magnitude increases, S/N$_{\text{PSD}}$ tends to decrease. This reflects the fact that the noise level in the TESS light curves increases for fainter objects. However, the S/N also depends on the amplitude of the astrophysical signal.
Individual polars exhibit different amplitudes in their periodicities due to factors such as the inclination of the system, which can affect the visibility of the hotspot. 
Therefore, the S/N$_{\text{PSD}}$ values present a significant dispersion for a given TESS magnitude. 
This means that $({\rm S/N_{PSD})_{min}}$ does not translate to a  definite value for the limiting TESS magnitude for a detectable period, but only an approximate range, where prudence should rule the interpretation of eventual period detections. Specifically, from the 223 observed light curves in which  S/N$_{\text{PSD}}$ could be estimated, approximately $9\%$ of the light curves with TESS magnitudes $T \gtrsim 17$\,mag have a S/N$_{\text{PSD}}$ below the threshold of $({\rm S/N_{PSD}})_{\rm min} = 0.004$, suggesting a false positive rate of around $9\%$ for systems fainter than $T \gtrsim 17$\,mag.
For such faint systems, the actual S/N$_{\text{PSD}}$ associated with a period measurement should be calculated following Appendix~\ref{sec:SNR} to assess the reliability of the measurement. For TESS magnitudes $T \gtrsim 19$\,mag, not covered by the sample of observations of this work, the same trend observed in Fig.~\ref{Fig.SNR_vs_Magnitude} is expected to apply and the percentage of light curves with S/N$_{\text{PSD}}$ below the detection reliability threshold likely increases. 

\section{Summary and conclusions}
\label{Summary}

We determined  orbital periods for $93$ out of 95 polars with TESS two-minute cadence light curves from Sectors\,1-63. For $85$ systems our measured values are consistent with values previously reported in the literature. The remaining ten objects comprise four asynchronous CVs with complex power-spectra and four systems where the TESS-based periods differ from those reported in the literature. For only 2 of the $95$ systems no period could be found because of high noise levels in their TESS light curves.   

For the four asynchronous systems our detailed analysis yielded 
an identification of both the orbital period, $\text{P}_{\text{orb}}$, and the spin period of the primary, $\text{P}_{\text{spin}}$ (see Table~\ref{table:asynchronous}). We detected new side-band periods for BY\,Cam and IGR\,J19552+0044, and revised the multiple periods present in the TESS light curves of CD\,Ind, confirming the physical origin of many of the low-power side-band periods presented in the previous literature (see Appendix~\ref{sec:periods_asynchronous}). 

A detailed examination of the four polars where our period determination is discrepant from the previous literature values yielded revised or improved values. Some of these measurements, however, remain tentative due to a low S/N and follow-up observations are recommended.

\subsection{Reliability of TESS-based CV period detections}\label{subsect:conclusion_reliability}

Earlier studies have highlighted the potential of TESS to refine or  improve period detections of CVs (see \citealt{Bruch_5}). 
The overall good agreement between our  measured orbital periods and the literature values for the majority of the polars  
confirms the suitability and efficiency of TESS light curves for the measurement of orbital periods in such systems. Moreover, it 
underscores the robustness of our  analysis methodology.  

Specifically, by simulating periodic light curves with varying noise levels and testing whether the orbital period can be recovered, we established  a probabilistic framework for the detection success across different S/N$_{\text{PSD}}$ intervals, where the S/N$_{\text{PSD}}$ of a light curve is a measure of the significance of the frequency peak associated with the orbital period with respect to the noise. A threshold value of $({\rm S/N}_{\text{PSD}})_{\rm min} = 0.004$ was determined at which the detection probability of orbital periods in TESS light curves of polars sharply decreases from a nearly $100\%$ recovery rate to below the $50\%$ level (see Fig.~\ref{Fig.Prob_vs_SNR}). 
The comparison of the measured ${\rm S/N_{ PSD}}$ in an observed TESS light curve with  this threshold enables us to estimate the reliability of a period detection. 

We note that the above ${\rm S/N_{\rm PSD}}$ threshold value was derived from simulations based on a specific observed TESS light curve—TIC\,840418301 (sector\,48) with an orbital period of $\text{P}_{\text{orb}}=82.4$\,min—and could be subject to some variation for systems with longer periods and when red noise significantly contributes to the power-law shape of the PSD. However, in CVs, the  orbital periods fall within a confined region of high frequencies in the PSD of TESS light curves. Since red noise affects lower-frequency power, its consideration would only lead to a more optimistic ${\rm S/N_{\rm PSD}}$ threshold, i.e. a lower limiting ${\rm S/N_{\rm PSD}}$ value (see Eq.~\ref{eq:noise_power} in Appendix~\ref{sec:SNR}), and it would not obscure the frequency peak associated with the orbital period (see Fig.~\ref{Fig.SNR_psd} in Appendix~\ref{sec:SNR}).

The S/N$_{\text{PSD}}$ was measured in all observed TESS light curves of our sample. 
We found that for systems with TESS magnitudes $T \gtrsim T_{\rm lim} \approx 17$\,mag, a significant fraction of light curves were found to be too noisy to accurately detect the orbital period. We note that the value for $T_{\rm lim}$ is not a precise limit because 
polars are variable, and they can exhibit both high and low states during an observation. Transitions between these states occur on timescales of days to years, with brightness changes of a few magnitudes (see \citealt{long_term_variations}). 
In addition, \cite{short_term_variations} found that short-duration accretion states are relatively common among polars. These states can last from a few days  to $\sim50$\,days.
Based on the trend observed in Fig.~\ref{Fig.SNR_vs_Magnitude}, we suggest the following practical guideline for detection reliability based on TESS magnitude:
\begin{itemize}
\item Reliable: $T < 17$\,mag;
\item Uncertain: $17 < T < 19$\,mag;
\item Likely unreliable: $T > 19$\,mag. 
\end{itemize} 
In the future we intend to complement this study with samples covering fainter TESS magnitudes in order to find the approximate limiting brightness that completely impedes the detection of orbital periods in CVs.
Our methodology will also be useful for the confirmation of CV candidates and, more generally, in the search for periods with TESS for other types of objects. 

\subsection{Noise level in TESS light curves}\label{subsect:conclusion_sflat}

We have expanded on our earlier studies of the dependence of the noise level in TESS light curves on the brightness of the stars. Specifically, we have determined the correlation between the TESS magnitude and the noise level in flattened TESS light curves, that had been cleaned from the known astrophysical signals (see Fig.~\ref{Fig.Sflat_vs_Tmag}). Compared to the samples of M dwarfs for which the same relation was   studied by us (\citealt{Sflat_1}, \citealt{Sflat_2}), the polars have much fainter $T$ magnitudes. To bridge the gap to the polars, we complement   the bright and nearby M dwarfs from our earlier works with a fainter M dwarf sample. Across the M dwarfs  we observe a well defined lower envelope and a sparse population of stars with higher noise levels that—according to  our earlier studies—is likely associated with intrinsic stellar variability hidden in the noise. The polars extend this trend of increasing noise with $T$ magnitude to fainter stars, albeit with a larger scatter. We suspect that the scatter of the noise in polars for given $T$ magnitude derives from flickering and from the fact that $T$ magnitude may not represent the actual brightness of the polar at the time of observation, as discussed in Sect.~\ref{subsect:conclusion_reliability}.
A dedicated study would be useful to constrain the origin of the noise level in TESS light curves of polars. 

We provide a functional form for the lower envelope of the noise in TESS light curves, expressed through the parameter $S_{\rm flat}$.
This relation allows   the expected intrinsic noise in TESS light curves to be estimated based on the object's brightness.  High noise levels can have a significant impact on the detection of low-amplitude astrophysical modulations in TESS light curves, and by characterising this noise we provide a useful reference for future variability studies in TESS light curves. 

\subsection{TESS-polars in the \textit{Gaia} colour-magnitude diagram}
\label{subsect:conclusion_gaiacmd}

\begin{figure}
\centering
\includegraphics[width=0.95\linewidth]{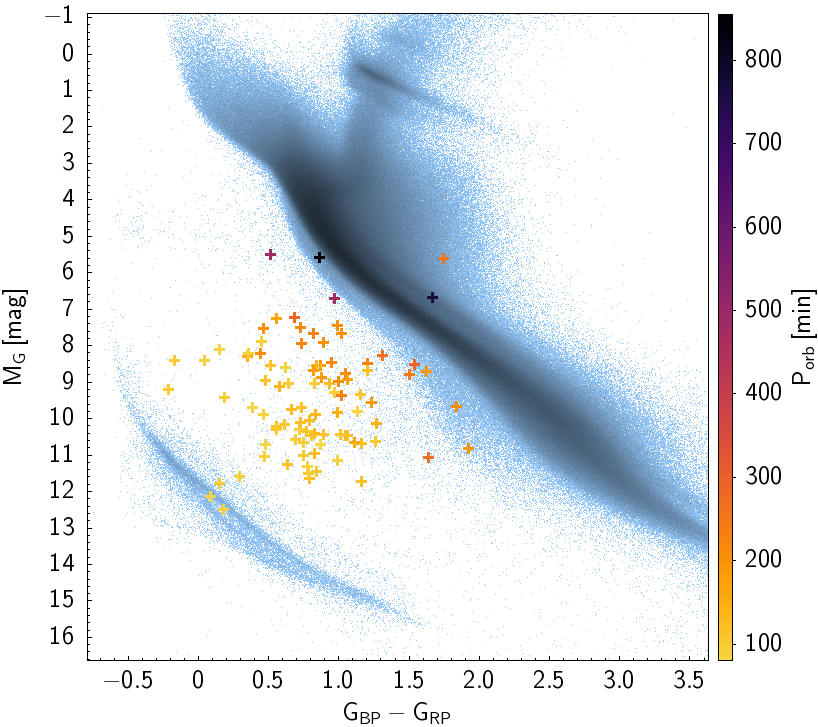}
\caption{\textit{Gaia} colour-magnitude diagram showing the sample of polars with TESS two-minute cadence light curves on a background of {\it Gaia}\,DR3 sources with distances from  \cite{Bailer-Jones}
and a parallax error of $<1\,\%$ of the parallax value. The colour-coding corresponds to the measured orbital periods.}
\label{Fig.CMD}
\end{figure}

Figure~\ref{Fig.CMD} displays the TESS sample of polars studied in this work in the {\it Gaia} colour-magnitude diagram (CMD). As can be seen from the comparison with  {\it Gaia}\,DR3 sources from \cite{Bailer-Jones}, polars are located between the main sequence and the white dwarf sequence.  
All the systems in our sample have a {\it Gaia} counterpart, and the absolute magnitude in the {\it Gaia} $G$ band was calculated making use of the \cite{Bailer-Jones} distances.
The colour-coding in Fig.~\ref{Fig.CMD} marks the measured orbital periods and shows that systems with longer periods are located closer to the main sequence, while systems with shorter periods 
approach the WD sequence. 
The long-term evolution of CVs is characterised by the transition from longer to shorter orbital periods due to the loss of angular momentum 
(see e.g. \citealt{Knigge_2} and \citealt{Kalomeni_1}). In less evolved systems, the donor star is still massive and bright such that its contribution dominates the luminosity of the system. As the systems evolve, the donor mass decreases as a consequence of the mass transfer to the WD and its contribution to the luminosity of the system becomes less significant. The observed period distribution is therefore in agreement with the expected evolution of CVs from the lower envelope of the main sequence to the WD sequence. This result is consistent with earlier findings reported by \cite{MCD_Gaia} from the CV catalogues from  \cite{Downes_2001} and \cite{203}. 
We note in Fig.~\ref{Fig.CMD} an apparent cavity around $M_{\rm G} \sim 10$ and $G_{\rm BP}-G_{\rm RP} \sim 0.5$. This feature arises due to the tracks that CVs follow in the Hertzsprung--Russell diagram as their mass accretion rate varies (see \citealt{Guillaume}). 
A measurement of donor masses will ultimately confirm this evolutionary sequence of CVs in the HR diagram. 

\section*{Data availability}
\label{sec:Data_availability}

Table~\ref{table:Master} is only available in electronic form at the CDS via anonymous ftp to \href{http://cdsarc.u-strasbg.fr/}{cdsarc.u-strasbg.fr} (\href{ftp://130.79.128.5/}{130.79.128.5}) or via 
\url{http://cdsweb.u-strasbg.fr/cgi-bin/qcat?J/A+A/}.

\begin{acknowledgements}
We wish to thank the anonymous referee for constructive comments and suggestions that helped improve the clarity and quality of this article. We are also grateful to Alex Binks and Simone Scaringi for helpful discussions and valuable comments on this work. Santiago Hernández-Díaz acknowledges financial support from Deutsche Forschungsgemeinschaft (DFG) under grant number STE\,1068/6-2. This work is based on data collected by the TESS mission and obtained from the MAST data archive at the Space Telescope Science Institute (STScI). Funding for the missions is provided by the NASA Explorer Program and the NASA Science Mission Directorate. STScI is operated by the Association of Universities for Research in Astronomy, Inc., under NASA contract NAS 5-26555. This work has made use of data from the European Space Agency (ESA) mission \textit{Gaia} (\href{https://www.cosmos.esa.int/gaia}{https://www.cosmos.esa.int/gaia}), processed by the \textit{Gaia} Data Processing and Analysis Consortium (DPAC, \href{https://www.cosmos.esa.int/web/gaia/dpac/consortium}{https://www.cosmos.esa.int/web/gaia/dpac/consortium}). Funding for the DPAC has been provided by national institutions, in particular the institutions participating in the \textit{Gaia} Multilateral Agreement.
\end{acknowledgements}

\bibliography{BIB.bib}

\begin{appendix}





\onecolumn
\begin{table*}[h!]
\section{Sample properties and results from period search for the catalogue of polars}
\label{sec:Table}

Table~\ref{table:Master} provides a summary of the structure of the main data table resulting from our work. It comprises basic coordinates, identifiers, photometry from TESS and {\it Gaia}, and the results obtained from the period search on their TESS light curves.

\caption {Description of the 24 columns in the sample of polars, including identifiers, coordinates, photometric data, orbital period measurements, and references. \newline}
\label{table:Master} 
\centering
\begin{tabular}{clll}
\hline\hline             
\# & Name & Unit & Description \\
\hline
1 & System &  & Object's common name. \\
2 & GaiaDR3 &  & \textit{Gaia} ID from data release 3. \\
3 & ra\_DR3 & deg & \textit{Gaia}\,DR3 Right Ascension. \\
4 & dec\_DR3 & deg & \textit{Gaia}\,DR3 Declination. \\
5 & TIC\_ID &  & TESS Input Catalogue ID. \\
6 & Sectors &  & Observed TESS sectors. \\
7 & T\_mag & mag & TESS magnitude. \\
8 & e\_T\_mag & mag & Uncertainty in TESS magnitude. \\
9 & bp\_phot & mag &  \textit{Gaia} BP mean magnitude. \\
10 & rp\_phot & mag & \textit{Gaia} RP mean magnitude. \\
11 & g\_phot & mag & \textit{Gaia} G mean magnitude. \\
12 & Porb\_LS & min & Lomb-Scargle sector-average orbital period. \\
13 & e\_Porb\_LS & min & Uncertainty in Porb\_LS.  \\
14 & Porb\_ACF & min & Autocorrelation function sector-average orbital period. \\
15 & e\_Porb\_ACF & min & Uncertainty in Porb\_ACF. \\
16 & Porb\_SINE & min & Sine fitting sector-average orbital period. \\
17 & e\_Porb\_SINE & min & Uncertainty in  Porb\_SINE.  \\
18 & Porb\_FS & min & Fourier power spectrum sector-average orbital period.  \\
19 & e\_Porb\_FS & min & Uncertainty in Porb\_FS. \\
20 & Porb & min & Final adopted orbital period. \\
21 & e\_Porb & min & Uncertainty in final Porb. \\
22 & Corr\_Flag &  & $1/2 \,\text{P}_{\text{orb}}$ correction flag. \\
23 & Porb\_lit & min & Literature-reported orbital period. \\
24 & Reference &  & Literature source for Porb\_lit. \\
\hline
\end{tabular}
\end{table*}

\section{Periodicities detected in asynchronous systems}
\label{sec:periods_asynchronous}

\begin{table*}[h!]
    \caption{Identified periods in the TESS light curves of TIC\,339374052 (alias BY\,Cam) (sectors\,19 and 59).}
    \label{table:BYCam}
    \centering
    \begin{tabular}{c c c c l}
        \hline\hline
        ID & Period (min) & S/N$_{\text{PSD, 19}}$ & S/N$_{\text{PSD, 59}}$ &  Interpretation \\
        \hline
        P$_{1}$  & $98.2546\pm0.0042$  & 0.015 &  & $2\pi \left(4\omega-2\Omega\right)^{-1}$ \\
        P$_{2}$  &  $99.694\pm 0.017$ & 0.18 & 0.031  &  $2\pi \left(2\omega\right)^{-1}$ \\
        P$_{3}$  &  $100.626\pm 0.087$ & 0.093 & 0.045  &  $2\pi \left(2\Omega\right)^{-1}$ \\
        P$_{4}$  &  $101.3114\pm0.0045$ & 0.022 &  &  $2\pi \left(3\Omega-\omega\right)^{-1}$ \\
        P$_{5}$  &  $102.3134\pm0.0047$ & 0.023 &  &  $2\pi \left(5\Omega-3\omega\right)^{-1}$ \\
        P$_{6}$  &  $191.4018\pm0.0074$ & 0.016 &  &  $2\pi \left(5\omega-4\Omega\right)^{-1}$ \\
        P$_{7}$  & $193.757\pm0.066$  & 0.036 & 0.018 &  $2\pi \left(4\omega-3\Omega\right)^{-1}$ \\
        P$_{8}$  & $195.894\pm 0.027$ & 0.047 & 0.069 &  $2\pi \left(3\omega-2\Omega\right)^{-1}$ \\
        P$_{9}$  &  $197.53\pm0.20$ & 0.92 & 1.52 &  $2\pi \left(2\omega-\Omega\right)^{-1}$ \\
        P$_{10}$ & $199.08\pm 0.24$  & 0.057 & 0.090 &  $2\pi \left(\omega\right)^{-1}$ \\
        P$_{11}$ & $201.220\pm 0.072$  & 0.043 & 0.13 &  $2\pi \left(\Omega\right)^{-1}$ \\
        P$_{12}$ & $202.8847\pm0.0083$ & 0.050 &  &  $2\pi \left(2\Omega-\omega\right)^{-1}$ \\
        P$_{13}$ & $205.29\pm0.94$ & 0.033 & 0.047 &  $2\pi \left(3\Omega-2\omega\right)^{-1}$ \\
        P$_{14}$ & $2312\pm 450$  & 0.052 & 0.068 &  $2\pi \left(6\omega-6\Omega\right)^{-1}$ \\
        P$_{15}$ &  $3336\pm 113$ & 0.038 & 0.14 &  $2\pi \left(5\omega-5\Omega\right)^{-1}$ \\
        P$_{16}$ &  $4360\pm273$ & 0.065 & 0.13 &  $2\pi \left(4\omega-4\Omega\right)^{-1}$ \\
        P$_{17}$ & $5340\pm241$  & 0.72 & 0.21 &  $2\pi \left(3\omega-3\Omega\right)^{-1}$ \\
        P$_{18}$ & $6827\pm43$  & 0.075 &  &  $2\pi \left(2\omega-2\Omega\right)^{-1}$ \\
        P$_{19}$ & $11359\pm37$  &  & 0.38 &  $2\pi \left(\omega-\Omega\right)^{-1}$ \\
        \hline
    \end{tabular}
\end{table*}

\begin{table*}[h!]
    \caption{Identified periods in the TESS light curve of TIC\,228975750 (alias IGR\,J19552+0044) (sector\,54).}
    \label{table:IGRJ19552}
    \centering
    \begin{tabular}{c c c l}
        \hline\hline
        ID & Period (min) &  S/N$_{\text{PSD}}$ & Interpretation \\
        \hline
        P$_{1}$  & $40.6402\pm0.0034$  & 0.011 & $2\pi \left(2\omega\right)^{-1}$ \\
        P$_{2}$  & $41.2167\pm0.0033$  & 0.014  & $2\pi \left(2\Omega\right)^{-1}$ \\
        P$_{3}$  & $77.0539\pm0.0046$  &  0.051 & $2\pi \left(4\omega-3\Omega\right)^{-1}$ \\
        P$_{4}$  & $81.3011\pm0.0015$  &  0.39 & $2\pi \left(\omega\right)^{-1}$ \\
        P$_{5}$  & $83.0153\pm0.0086$  &  0.012 & $2\pi \left(\Omega\right)^{-1}$ \\
        P$_{6}$  & $86.041\pm0.014$  &  0.015  & $2\pi \left(3\Omega-2\omega\right)^{-1}$ \\
        \hline
    \end{tabular}
\end{table*}

\begin{table*}[h!]
    \caption{Identified periods in the TESS light curves of TIC\,231666244 (alias CD\,Ind) (sectors\,1 and 27).}
    \label{table:CDInd}
    \centering
    \begin{tabular}{c c c c l}
        \hline\hline
        ID & Period (min) & S/N$_{\text{PSD, 1}}$ & S/N$_{\text{PSD, 27}}$ & Interpretation \\
        \hline
        P$_{1}$  & $36.6782 \pm 0.0019$  & 0.034  &  0.062  & $2\pi \left(5\omega-2\Omega\right)^{-1}$ \\
        P$_{2}$  & $54.5517\pm0.0016$    &  0.018 &    & $2\pi \left(5\omega-3\Omega\right)^{-1}$ \\
        P$_{3}$  & $54.8206 \pm 0.0013$    & 0.042  &  0.24  & $2\pi \left(4\omega-2\Omega\right)^{-1}$ \\
        P$_{4}$  & $55.111 \pm 0.018$    &  0.14 &  0.046  & $2\pi \left(3\omega-\Omega\right)^{-1}$ \\
        P$_{5}$  & $55.4101\pm0.0016$    &  0.11 &    & $2\pi \left(2\omega\right)^{-1}$ \\
        P$_{6}$  & $106.35 \pm 0.12$    &  0.013 &  0.048  & $2\pi \left(5\omega-4\Omega\right)^{-1}$ \\
        P$_{7}$  & $107.31 \pm 0.19$    &  0.055 &  0.036  & $2\pi \left(4\omega-3\Omega\right)^{-1}$ \\
        P$_{8}$  & $108.448 \pm 0.044$   &  0.048 &  0.042   & $2\pi \left(3\omega-2\Omega\right)^{-1}$ \\
        P$_{9}$  & $109.657 \pm 0.017$    &  0.86 &  2.25  & $2\pi \left(2\omega-\Omega\right)^{-1}$ \\
        P$_{10}$ & $110.9200\pm 0.0023$    & 0.19  &    & $2\pi \left(\omega\right)^{-1}$ \\
        P$_{11}$ & $111.9719\pm0.0024$    &  0.22 &    & $2\pi \left(\Omega\right)^{-1}$ \\
        P$_{12}$ & $113.1220\pm 0.0014$    &   & 0.064   & $2\pi \left(2\Omega-\omega\right)^{-1}$ \\
        P$_{13}$ & $2114\pm 44$    &  0.058 &  0.087  & $2\pi \left(5\omega-5\Omega\right)^{-1}$ \\
        P$_{14}$ & $2653\pm25$    &  0.091 &  0.066  & $2\pi \left(4\omega-4\Omega\right)^{-1}$ \\
        P$_{15}$ & $3472\pm133$    & 0.10  &  0.078  & $2\pi \left(3\omega-3\Omega\right)^{-1}$ \\
        P$_{16}$ & $5202\pm139$    &  0.59 &  0.065  & $2\pi \left(2\omega-2\Omega\right)^{-1}$ \\
        P$_{17}$ & $11147\pm 68$    &  0.10 &    & $2\pi \left(\omega-\Omega\right)^{-1}$ \\
        \hline
    \end{tabular}
\end{table*}

\begin{table*}[h!]
    \caption{Identified periods in the TESS light curve of TIC\,369210348 (alias Paloma) (sector\,19).}
    \label{table:Paloma}
    \centering
    \begin{tabular}{c c c l}
        \hline\hline
        ID & Period (min) & S/N$_{\text{PSD}}$ & Interpretation \\ 
        \hline
        P$_{1}$ & $68.1146\pm0.0034$  & 0.088 & $2\pi \left(2\omega\right)^{-1}$ \\  
        P$_{2}$ & $72.9782\pm0.0059$  & 0.041 & $2\pi (\omega + \Omega)^{-1}$ \\  
        P$_{3}$ & $78.5897\pm0.0077$  & 0.026 & $2\pi \left(2\Omega\right)^{-1}$ \\  
        P$_{4}$ & $136.235\pm0.025$ & 0.026 & $2\pi \left(\omega\right)^{-1}$ \\  
        P$_{5}$ & $157.187\pm0.018$ &  0.093 & $2\pi \left(\Omega\right)^{-1}$ \\  
        P$_{6}$ & $1022.7\pm1.2$ & 0.055 & $2\pi (\omega - \Omega)^{-1}$ \\  
        \hline
    \end{tabular}
\end{table*}

\onecolumn
\section{Simulation of TESS light curves}
\label{sec:simulation}

\begin{figure}[h!]
    \centering
    \includegraphics[width=0.7\linewidth]{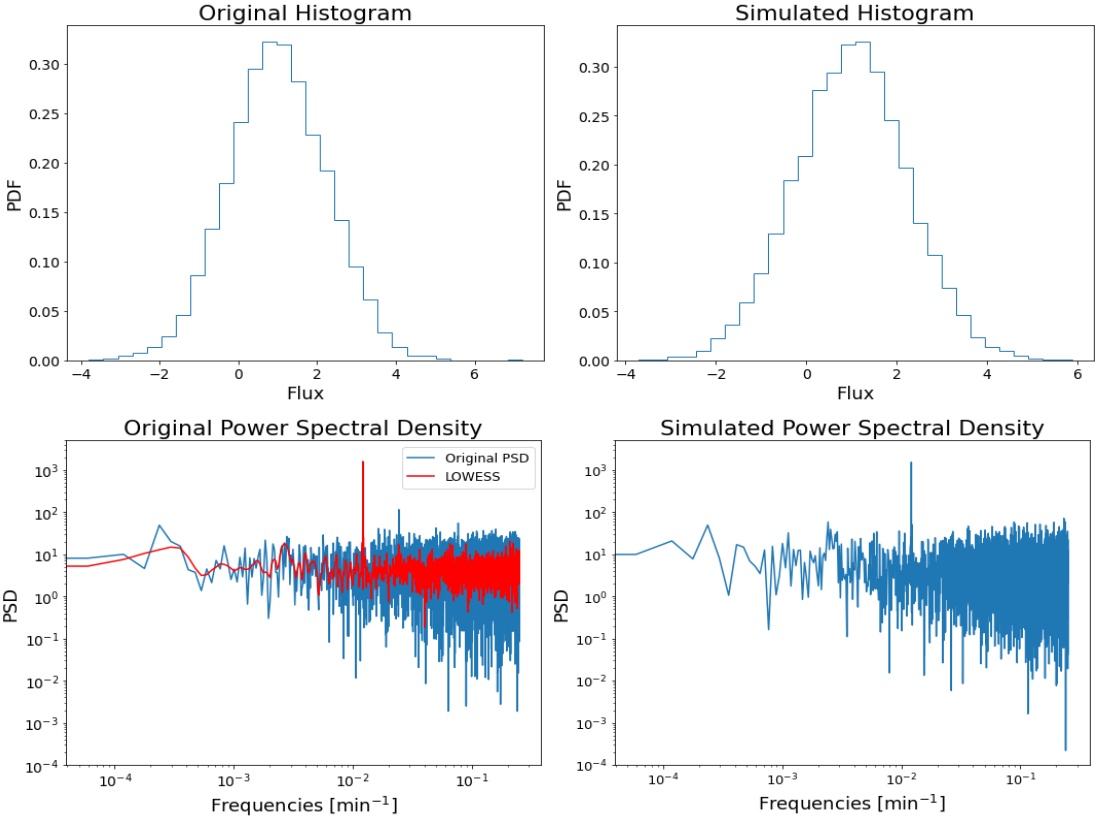}
    \caption{PDF and PSD of the original observed TESS light curve of TIC\,840418301 (sector\,48) and a simulated light curve for comparison. The normal distribution characterising the PDF is conserved and the simulated PSD correctly replicates the PSD of the observed light curve.}
    \label{Fig.Simulated_Light_Curve}
\end{figure}

The algorithm from Timmer and König (\citealt{TK95}) (hereafter denoted as TK95 algorithm), has been implemented to simulate TESS light curves. This method generates synthetic light curves from an underlying PSD model. In this study, the PSD model corresponds to a smoothed version of the observed PSD from TIC\,840418301 (sector\,48). 

To compute the PSD of the observed light curve of TIC\,840418301 (sector\,48), a Kaiser window with $\beta=5$ is used. This value for the parameter $\beta$ is higher than the value used in the generation of the Fourier power spectrum for the detection of the orbital periods, where $\beta=3$ was used. In the detection of orbital periods, maintaining a high-frequency resolution was primary. On the contrary, in the simulation of light curves, a greater compromise with the spectral leakage reduction is preferred, effectively diminishing the distortion in the PSD estimation. However, a precise reproduction of the orbital period signal in the simulated light curves is required, and therefore a low value for $\beta$ is still convenient.

To estimate the underlying PSD of the observed light curve of TIC\,840418301 (sector\,48), the LOWESS (Locally Weighted Scatterplot Smoothing) non-parametric regression method (\citealt{LOWESS_1}, \citealt{LOWESS_2}) is additionally implemented. The LOWESS method   reveals the underlying form of the PSD, smoothing the high-frequency noise, which is enhanced during the simulation process, and capturing the local variations represented by the frequency peaks. To this extent, the non-parametric nature of the method solves the problem of specifying a global analytical function to fit a model to the PSD of the observed light curve. This estimated PSD will be the target PSD which should be matched by the TK95 algorithm in the production of simulated light curves.

The TK95 algorithm generates a stochastic realisation of a PSD model. In our case, this means that the height of the frequency peak can differ across different simulations. To ensure consistency between the frequency peak height in the observed PSD of TIC\,840418301 (sector\,48) and that of simulated light curves, only simulations where the peak height deviated by no more than 5\% from the observed value were considered.

The TK95 algorithm is designed to preserve only the initial two statistical moments of
the original data set, the mean value and the variance. Higher-order statistical moments,
such as skewness and kurtosis, are disregarded in the resulting simulated data. Thus, the
TK95 algorithm is only suitable for generating simulated data sets that exhibit a Gaussian
distribution. To produce simulations that match the observed light curve's statistical characteristics, the observed light curve's PDF must also be normally distributed. For the observed light curve of TIC\,840418301 (sector\,48), normality diagnostics confirm the light curve's normal distribution, making it suitable as a template for simulation-recovery tests (details in Appendix~\ref{sec:Normality}).

Another algorithm, based on TK95, can match both the PSD and PDF of observed light curves (\citealt{Emmanoulopoulos}). While advantageous for the simulation of observed light curves that present deviations from normality, it struggles to reproduce the frequency peaks in the PSD of CVs light curves, leading to its exclusion from this analysis.

Finally, this analysis involves generating simulated light curves with varying levels of white noise. The  white noise, $wn \sim \mathscr{N}(0, \sigma^{2}_{wn})$, is added to the simulated light curves, with its level regulated by the variance $\sigma^{2}_{wn}$.

\onecolumn
\section{Normality diagnostics}
\label{sec:Normality}

\begin{figure}[h!]
	\centering
	\includegraphics[width=0.8\linewidth]{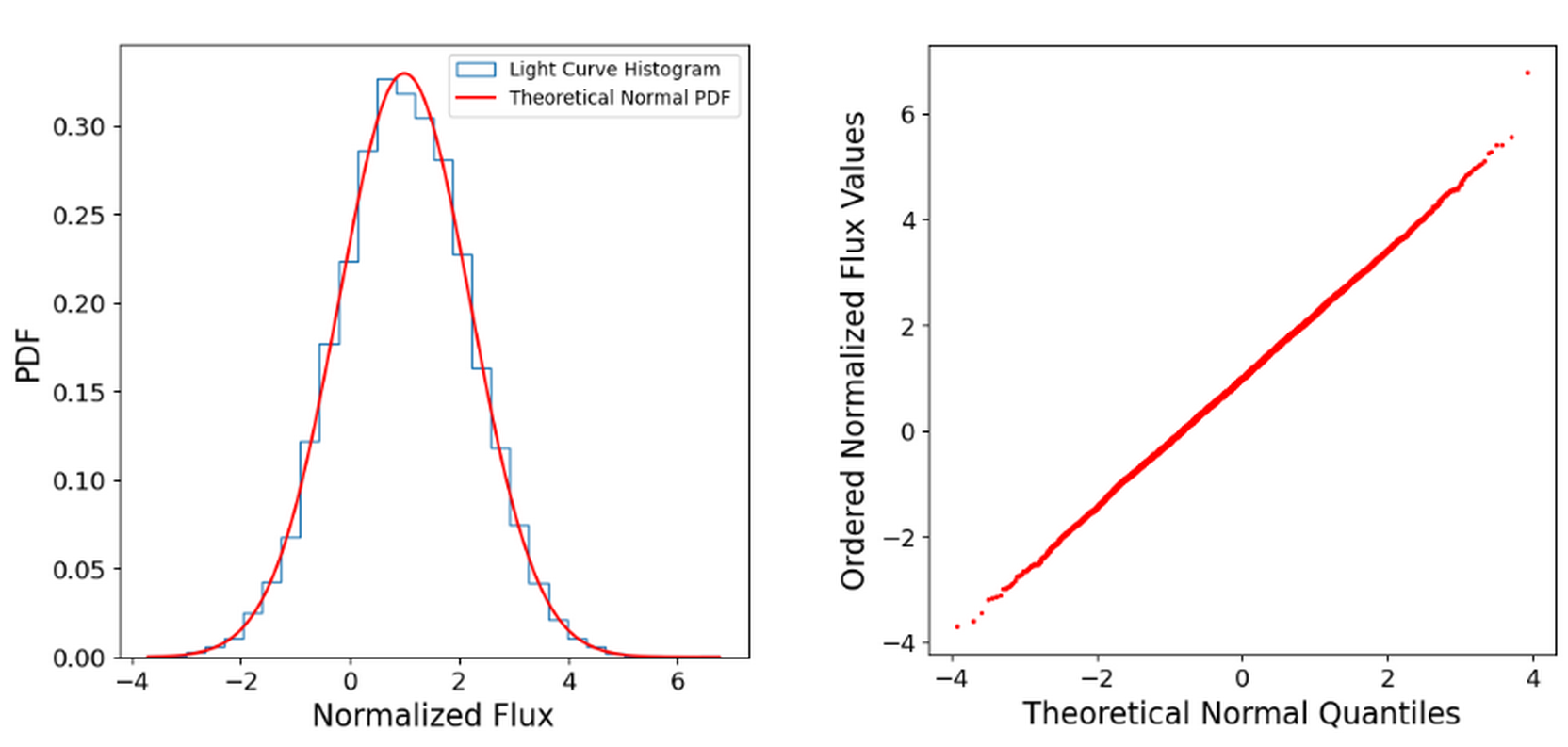}
	\caption{Normality plots for the light curve of TIC\,840418301 in sector\,48. Left: Light curve histogram with the theoretical normal probability density function overplotted. The histogram's values correspond with the probability density function at the bin. The histogram is normalised such that the integral over the entire range is 1. Right: Ordered sample data against the theoretical quantiles of the normal distribution.}
	\label{Fig:Normal_Plots}
\end{figure}

The normality of the observed light curve of TIC\,840418301 (sector\,48) was evaluated using both graphical and statistical methods.

As a first step, the light curve was flattened to subtract the periodic modulation, effectively isolating the noise. This ensures that the normality evaluation focuses solely on the stochastic noise, rather than the deterministic periodicity present in the original data.

In Fig.~\ref{Fig:Normal_Plots}, two graphical representations are presented: the light curve’s histogram (left panel) and a normal probability plot (right panel). The histogram closely matches the theoretical normal PDF defined by the light curve's mean and standard deviation. The normal probability plot reveals minor deviations from linearity at the extremes, indicating certain level of truncation in the data, that is, there are not observations covering the full range given by the normal distribution. Aside from this, no important deviations from normality are observed.

The skewness and kurtosis were estimated as $\sqrt{b_{1}}=0.005\pm 0.020$ and $b_{2}=3.043\pm 0.039$, respectively. The obtained skewness is, within the standard error, in complete agreement with a normal distribution, while the obtained kurtosis value is sufficiently close to 3 and deviations from normality can be considered as negligible.

The Shapiro-Wilk normality test (\citealt{Shapiro_Wilk}) was performed yielding a $\text{p-value}=0.67$.  The Shapiro-Wilk test presents one important defect to be taken into account: it does not perform well with large data sets. Hence, additional normality diagnostics are considered.

The D’Agostino’s K2 test (\citealt{Omnibus_test}) and the Anderson-Darling test (\citealt{Anderson-Darling}) were implemented producing a $\text{p-value}=0.52$ and a $A^{2}$ statistic of $A^{2}=0.355$, respectively. For a significance level of 0.05, the associated critical value in the Anderson-Darling test is 0.787.
 
At a typical significance level of 0.05, none of the normality tests rejects the null hypothesis that the light curve is drawn from a normal distribution.

In this work, the Python implementations of the aforementioned normality tests included in the package \textsl{Scipy} were used.

\onecolumn
\section{Signal-to-noise ratio measurement}
\label{sec:SNR}

\begin{figure*}[h!]
	\centering
	\includegraphics[width=0.75\linewidth]{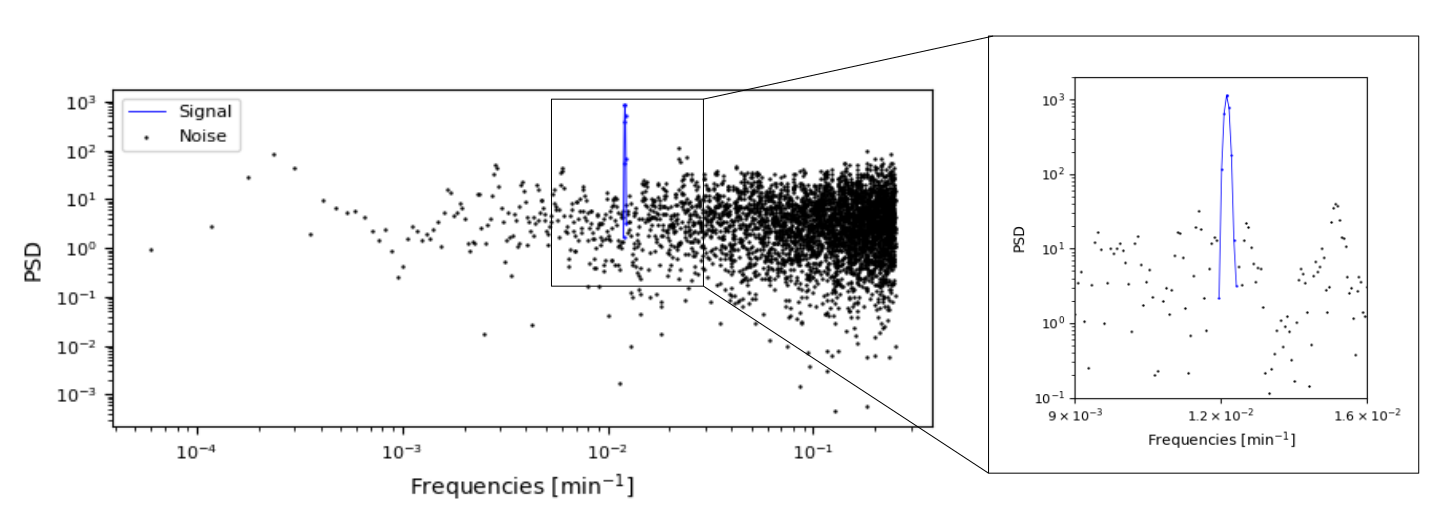}
	\caption{Measurement of S/N$_{\text{PSD}}$ in a simulated light curve. The frequency peak in the PSD of the light curve corresponding to the orbital period is identified as the signal. After the removal of the frequency peak, the remaining signal is associated with the noise.}
	\label{Fig.SNR_psd}
\end{figure*}

The measurement of the signal-to-noise ratio in the PSD, S/N$_{\text{PSD}}$, of simulated light curves is conducted with a methodology based on Fourier transforms. As explained in Sect.~\ref{subsect:Simulation-Recovery_Test}, the light curve of 
TIC\,840418301 (sector\,48) serves as a template for our simulations.

For each simulated light curve, we first computed the PSD with the use of a Kaiser window (\citealt{Kaiser_Window}) with $\beta =12$. The choice of a high value for the parameter $\beta$ responds to the priority of reducing the spectral leakage in the PSD, which would otherwise affect the signal's power estimation. For the template light curve of TIC\,840418301 (sector\,48), the precise location of the frequency peak associated with the orbital period is already known from the period search described in Sect.~\ref{Search_Period} and a loss in the frequency resolution does not act to the detriment of the S/N$_{\text{PSD}}$ measurement. The measured orbital period in TIC\,840418301 (sector\,48) is $\text{P}_{\text{orb}}=82.4$\,min. The frequency peak in the PSD that corresponds to the orbital period is therefore readily identified.  This peak represents the signal of interest and its associated power is computed by integrating the area under the frequency peak in the PSD,
\begin{equation}
\mathscr{P}_{\text{signal}} := \int_{f_{\text{lower}}}^{f_{\text{upper}}} PSD(f) df,
\end{equation}
\noindent where $f_{\text{lower}}$ and $f_{\text{upper}}$ denote the lower and upper frequencies that determine the range that captures the frequency peak in its entirety.

Next, the frequency peak corresponding with the orbital period is removed from the PSD. For the removal of the frequency peak, consecutive data points with continuously decreasing values on both sides of the peak's maximum are considered, that is, data points $PSD(f_{i-1})$ holding $PSD(f_{i-1})<PSD(f_{i})$ are selected on the left of the peak's maximum and data points $PSD(f_{i+1})$ holding $PSD(f_{i+1})<PSD(f_{i})$ are selected on the right of the peak's maximum, where $f_{i}$ denotes the discrete Fourier frequencies.
All other parts of the PSD are then associated with the noise (see Fig.~\ref{Fig.SNR_psd}) and its power is computed, as before, by integration,

\begin{equation}
\mathscr{P}_{\text{noise}} := \int_{f_{\text{min}}}^{f_{\text{max}}} PSD(f) df,
\label{eq:noise_power}
\end{equation}
\noindent where $f_{\text{min}}$ and $f_{\text{max}}$ denote the minimum and maximum frequencies in the PSD. 

The S/N$_{\text{PSD}}$ is thus
\begin{equation}
S/N_{\text{PSD}} := \dfrac{\mathscr{P}_{\text{signal}}}{\mathscr{P}_{\text{noise}}}.
\end{equation} 
\noindent 
We note that, in the measurement of the S/N$_{\text{PSD}}$, the DC component, given by the zero Fourier frequency component, has not been included, as the relevant information here considered is the predominance of the frequency peak associated with the orbital period with respect to the noise that, potentially, could hinder its detection.

\end{appendix}
\end{document}